\documentclass[12pt,a4paper]{article}
\pdfoutput=1
\usepackage{graphicx}
\usepackage[T1]{fontenc}
\usepackage[sc,osf]{mathpazo}
\usepackage{a4wide}  
\usepackage{latexsym,amsthm,amsfonts,amsmath,mathrsfs,amssymb}
\usepackage[unicode,implicit]{hyperref}
\hypersetup{%
  pdftitle    = {Supersymmetric solutions of the cosmological, gauged, C magic model}
  pdfkeywords = {Yang-Mills, gravity, Einstein-Yang-Mills, black hole,
    monopole, hair, supergravity, supersymmetry, gauging},
  pdfauthor   = {Samuele Chimento, Tomas Ortin, Alejandro Ruiperez},
  plainpages  = true,
  colorlinks  = true,
  citecolor   = blue,
  urlcolor    = red,
  linkcolor   = black
}
\newcommand{\hepth}[1]{{\tt
\href{http://www.arXiv.org/abs/hep-th/#1}{hep-th/#1}}}

\newcommand{\arxiv}[1]{{\tt arXiv:\href{http://www.arXiv.org/abs/#1}{#1}}}
\newcommand{\doi}[1]{{\tt DOI:\href{http://dx.doi.org/#1}{#1}}}

\makeatletter
\@addtoreset{equation}{section}
\makeatother

\DeclareMathOperator{\Tr}{Tr}

\begin{document}

\begin{flushright}
\small

IFT-UAM/CSIC-18-008\\
February 3\textsuperscript{rd}, 2018\\
\normalsize
\end{flushright}

\vspace{1cm}

\begin{center}
 
  {\large {\bf Supersymmetric solutions of the cosmological, gauged,
      $\mathbb{C}$ magic model}}
 
\vspace{1.5cm}

\renewcommand{\thefootnote}{\alph{footnote}}
{\sl Samuele Chimento}\footnote{E-mail: {\tt Samuele.Chimento [at] csic.es}},
{\sl Tom\'{a}s Ort\'{\i}n}\footnote{E-mail: {\tt Tomas.Ortin [at] csic.es}}
{\sl and Alejandro Ruip\'erez}\footnote{E-mail: {\tt alejandro.ruiperez [at] uam.es}},

\setcounter{footnote}{0}
\renewcommand{\thefootnote}{\arabic{footnote}}

\vspace{1.5cm}

{\it Instituto de F\'{\i}sica Te\'orica UAM/CSIC\\
C/ Nicol\'as Cabrera, 13--15,  C.U.~Cantoblanco, E-28049 Madrid, Spain}\\

\vspace{1.5cm}


{\bf Abstract}

\end{center}

\begin{quotation}
  We construct supersymmetric solutions of theories of gauged
  $\mathcal{N}=1,d=5$ supergravity coupled to vector multiplets with a
  $\mathrm{U}(1)_{\rm R}$ Abelian (Fayet-Iliopoulos) gauging and an independent
  SU$(2)$ gauging associated to an $\mathrm{SU}(2)$ isometry group of the Real
  Special scalar manifold. These theories provide minimal
  supersymmetrizations of 5-dimensional SU$(2)$ Einstein-Yang-Mills theories
  with negative cosmological constant. We consider a minimal model with these
  gauge groups and the ``magic model'' based on the Jordan algebra
  $\bf{J}_{3}^\mathbb{C}$ with gauge group
  $\mathrm{SU}(3)\times\mathrm{U}(1)_{\rm R}$, which is a consistent truncation of
  maximal $\mathrm{SO}(6)$-gauged supergravity in $d=5$ and whose solutions
  can be embedded in Type~IIB Superstring Theory. We find several solutions
  containing selfdual $\mathrm{SU}(2)$ instantons, some of which asymptote to
  AdS$_{5}$ and some of which are very small, supersymmetric, deformations of
  AdS$_{5}$. We also show how some of those solutions can be embedded in
  Romans' $\mathrm{SU}(2)\times\mathrm{U}(1)$-gauged half-maximal supergravity, which was
  obtained by Lu, Pope and Tran by compactification of the Type~IIB Superstring
  effective action. This provides another way of uplifting those solutions to
  10 dimensions.
\end{quotation}

\newpage
\pagestyle{plain}

\tableofcontents


\section{Introduction}

Over the last 25 years, since the first dilaton black-hole and $p$-brane
solutions were found, there has been a continuous effort in finding and
studying solutions of supergravity theories in diverse dimensions, specially
if the supergravity theories describe the low-energy effective field theory
limit of a superstring theory. This continuous effort has been rewarded with
the discovery of many interesting solutions, some of which have revolutionized
the field. 

To a large extent, however, solutions with non-trivial non-Abelian fields
have been left out of this research effort. This was probably due to
several different reasons: the vast number of interesting Abelian solutions
one could work with, the expected complexity of the non-Abelian ones (all the
solutions of Einstein-Yang-Mills (EYM) theories which are not Abelian
embeddings are only known numerically), the expected violation of the no-hair
theorems in non-Abelian black holes, the loss of nice properties such as the
\textit{attractor mechanism} in extremal black holes
\cite{Ferrara:1995ih,Strominger:1996kf,Ferrara:1996dd,Ferrara:1996um,Ferrara:1997tw},
and our general lack of understanding of this kind of solutions.

From our viewpoint, the only way to increase our knowledge on the properties
of solutions (black holes, black strings, solitons...) with non-trivial,
non-Abelian Yang-Mills fields (non-Abelian solutions, in short), is to find
first many more. Fortunately, although this task may look extremely difficult
\textit{a priori}, it turns out that, just as in the Abelian case, one can use
supersymmetry to derive very powerful solution-generating
techniques. Typically, these techniques reduce the problem of finding
solutions of supergravity theories to the problem of solving a reduced number
of differential equations for functions, 1-forms etc. that play the r\^ole of
building blocks of the full solutions.\footnote{A complete review of these
  techniques with many references can be found in Ref.~\cite{Ortin:2015hya}.}

Thus, using the solution-generating techniques derived in
Refs.~\cite{Hubscher:2008yz,Bellorin:2007yp}, a large number of
asymptotically-flat non-Abelian solutions of different kinds (black holes,
strings and rings, global monopoles and instantons, multi-center black-hole
solutions, microstate geoemtries etc.) have been constructed over the last few
years in 4- and 5-dimensional non-Abelian-gauged supergravities with 8
supercharges which can be called Super-Einstein-Yang-Mills (SEYM) theories
because they are minimal supersymmetrizations of the EYM theories
\cite{Meessen:2008kb,Hubscher:2008yz,Bueno:2014mea,Meessen:2015nla,Meessen:2015enl,Ortin:2016bnl,Ramirez:2016tqc,Cano:2017sqy,Cano:2017qrq,Meessen:2017rwm,Avila:2017pwi}.
All these solutions were obtained in fully analytic form, which allows
showing, for instance, how the attractor mechanism works in a covariant
fashion in the non-Abelian context \cite{Huebscher:2007hj} and, more recently,
how a puzzle involving an apparent violation of the no-hair conjecture is
solved when the integration constants of the solution are expressed correctly
in terms of the charges of string-theory objects \cite{Cano:2017qrq}.

Extending this work to the asymptotically-AdS case requires important
modifications of the gaugings considered because the scalar potentials that
arise in the simplest gauging of non-Abelian isometries of the scalar
manifolds are, necessarily, either positive-definite (in the $d=4$ case) or
identically zero (in the $d=5$ case). The only way to produce the scalar
potential needed is to gauge a subgroup of the R-symmetry group
($\mathrm{U}(2)_{\rm R}$ in $d=4$ and $\mathrm{SU}(2)_{\rm R}$ in $d=5$) via
the introduction of Fayet-Iliopoulos (FI) terms. Both in the $d=4$ and $d=5$
cases the FI terms can be used to gauge either a $\mathrm{U}(1)_{\rm R}$ or a
$\mathrm{SU}(2)_{\rm R}$ subgroup. The Abelian $\mathrm{U}(1)_{\rm R}$ has
been studied extensively, but always in absence of any other non-Abelian
gauging. The non-Abelian $\mathrm{SU}(2)_{\rm R}$ case has been studied in
Refs.~\cite{Cariglia:2004kk,Ortin:2016sih} and turns out to be, technically,
much more complicated because the gauging of $\mathrm{SU}(2)_{\rm R}$ requires
the simultaneous gauging of a $\mathrm{SU}(2)$ subgroup of the isometry group
of the Special-K\"ahler ($d=4$) or Real-Special ($d=5$) scalar manifold. In
contrast, the Abelian $\mathrm{U}(1)_{\rm R}$ gauging never involves the
gauging of a single $\mathrm{U}(1)$ isometry in these theories.\footnote{These
  Abelian gaugings are, actually, not possible in these theories.}

In this paper we work in the framework of the 5-dimensional theories
($\mathcal{N}=1,d=5$ supergravity coupled to vector multiplets) and we are
going to consider the first of these possibilities: an Abelian
$\mathrm{U}(1)_{\rm R}$ gauging that will produce a scalar potential with AdS
vacua and, at the same time, an independent non-Abelian gauging of a subgroup
of the isometry group of the scalar manifold. The resulting theories can be
understood as the natural non-Abelian extension of those with an Abelian
gauging and additional vector multiplets. They can also be thought of as the
simplest supersymmetrization of the cosmological EYM theories (EYM plus a
cosmological constant). Thus, they may be expected to give us a handle in the
search for solutions of this system via the use of the supersymmetric
solution-generating techniques developed over the years in
Refs.~\cite{Gauntlett:2002nw,Gauntlett:2003fk,Gutowski:2004yv,Gauntlett:2004qy,Gutowski:2005id,Bellorin:2006yr,Bellorin:2007yp,Bellorin:2008we,Chimento:2016run,Chimento:2016mmd,Chimento:2017sen}. In
particular, we will be able to use the techniques of
Ref.~\cite{Chimento:2017tsv} to construct self-dual $\mathrm{SU}(2)$
instantons on K\"ahler spaces admitting a holomorphic isometry, which are one
of the main ingredients in those solution-generating techniques.

There are many possible models of $\mathcal{N}=1,d=5$ supergravity coupled to
vector supermultiplets and a good number of them admit the kind of gauging we
want to consider here. We have decided to consider, as a toy model, the
simplest of them admitting the gauging $\mathrm{SU}(2)\times
\mathrm{U}(1)_{\rm R}$ and, searching for a possible embedding of the
solutions in String Theory, the so-called ``$\mathbb{C}$ magic model'', that
admits a $\mathrm{SU}(3)\times \mathrm{U}(1)_{\rm R}$ gauging. In its ungauged
form, this model is one of the few $\mathcal{N}=1,d=5$ models that can be
obtained by consistent truncation of the maximal supergravity in $d=5$ and,
therefore, can be uplifted to any of the two maximal supergravities in $d=10$,
$\mathcal{N}=2A$ and $2B$ but, precisely with that gauging, it can be obtained
by a consistent truncation of the $\mathcal{SO}(6)$-gauged maximal
supergravity in $d=5$, which, in its turn, can be obtained by compactification
of the $\mathcal{N}=2B,d=10$ on S$^{5}$. Thus, in principle, all the solutions
of this theory, that we are going to call ``cosmological, gauged, $\mathbb{C}$
magic model'', are also solutions of $\mathcal{N}=2B,d=10$ supergravity, the
low-energy effective field theory of the Type~IIB Superstring, and, in
particular, the $\mathrm{AdS}_{5}$ vacuum of the cosmological, gauged,
$\mathbb{C}$ magic model corresponds to the maximally supersymmetric
$\mathrm{AdS}_{5}\times \mathrm{S}^{5}$ near-horizon limit of the D3-brane.

We have found that some solutions of the cosmological, gauged, $\mathbb{C}$
magic model can also be embedded in $\mathcal{N}=2B,d=10$ supergravity via the
$\mathrm{SU}(2)\times\mathrm{U}(1)$-gauge half-maximal supergravity obtained
by Romans \cite{Romans:1985ps} following the recipe given by
Lu, Pope and Tran in Ref.~\cite{Lu:1999bw}: there are two consistent
truncations (one of the cosmological, gauged, $\mathbb{C}$ magic model and
another of the gauged half-maximal supergravity) that lead to exactly the same
theory. This provides two different ways of uplifting these solutions to $\mathcal{N}=2B,d=10$ supergravity and an embedding into the
Type~IIB Superstring effective action to zeroth order in $\alpha'$.

This paper is organized as follows: in Section~\ref{sec-setup} we describe the
framework we are going to work in, introducing the formalism of gauged
$\mathcal{N}=1,d=5$ supergravities coupled to vector multiplets in
Section~\ref{sec-gaugedsugra} and the two particular models we are going to
consider in Sections~\ref{sec-simplemodel} and~\ref{sec-themodel}. In
Section~\ref{sec-susysolutions} we describe the general technique we use to
construct timelike supersymmetric solutions of generic gauged
$\mathcal{N}=1,d=5$ supergravities coupled to vector multiplets and, in
Section~\ref{sec-susysolutionscomogauged} we particularize this technique to
the kind of gaugings considered here. Then, in
Sections~\ref{sec-susysolutionssimplemodel}
and~\ref{sec-susysolutionsCmagicmodel} we apply the technique to the two
models we have chosen and construct the simplest solutions that have a
non-trivial non-Abelian field. Finally, in Section~\ref{sec-n4embedding} we
study the embedding of the solutions of the second model in the
$\mathrm{SU}(2)\times\mathrm{U}(1)$-gauge half-maximal supergravity showing in
Section~\ref{sec-truncated} the relation between the two consistent
truncations mentioned above.  Section~\ref{sec-conclusions} contains our
conclusions.

\section{The setup}
\label{sec-setup}

In this section we describe the two theories we are going to work with. They
are two different models of gauged $\mathcal{N}=1,d=5$ supergravity coupled to
vector supermultiplets with gauge groups consisting in a U$(1)$ factor
associated to a Fayet-Iliopopulos term and second, non-Abelian factor (SU$(2)$
and SU$(3)$) associated to the gauging of the isometry group of the (Real
Special) scalar manifold. $\mathcal{N}=1,d=5$ supergravity coupled to vector
supermultiplets with a non-Abelian gauging provides the minimal
supersymmetrization of 5-dimensional Einstein-Yang-Mills theory\footnote{The
  minimal, $\mathcal{N}=1$ supersymmetrization of a 5-dimensional
  Einstein-Yang-Mills (EYM) theory requires (apart from the introduction of
  fermions, which we set to zero here) the introduction of scalars to have
  complete vector supermultiplets. The scalars have to parametrize a Real
  Special manifold whose isometry group contains the gauge group. This may not
  be possible for arbitrary groups because, at the same time, the scalars must
  transform under the isometry group in a very precise way, which may demand
  the introduction of more vector fields. As we are going to see, the
  supersymmetrization of the SU$(3)$ EYM theory corresponds to a highly
  non-trivial ``magical model'' and has one extra vector field, the
  graviphoton. Besides the mere introduction of scalar fields through a
  $\sigma$-model, the supersymmetrized EYM theory (or Super-EYM (SEYM) theory)
  contains couplings between the scalar and vector fields and Chern-Simons
  terms for the vector fields which typically are absent in EYM theories. It
  is the contribution of the Chern-Simons terms gives that rise to very
  interesting and characteristic supersymmetric solutions of these theories.}.

Since the structure of these gaugings is somewhat complicated, but essential
to our goals, we start by reviewing gauged $\mathcal{N}=1,d=5$ supergravity
coupled to vector supermultiplets in general and, next, we describe in detail
the two models.

\subsection{Gauged $\mathcal{N}=1,d=5$ supergravity coupled to vector multiplets}
\label{sec-gaugedsugra}

A model of ungauged $\mathcal{N}=1,d=5$ supergravity coupled to $n$ vector
multiplets\footnote{Our conventions are those in
  Refs.~\cite{Bellorin:2006yr,Bellorin:2007yp} and the more recent
  Refs.~\cite{Meessen:2015enl,Chimento:2016mmd} which are those of
  Ref.~\cite{Bergshoeff:2004kh} with minor modifications and adaptations.} is
fully characterized by the constant, completely symmetric tensor $C_{IJK}$,
$I,J,\ldots=0,1,\cdots,n$ and its bosonic content is: the spacetime metric
$g_{\mu\nu}$, $n+1$ vector fields $A^{I}{}_{\mu}$ and $n$ scalars $\phi^{x}$,
$x,y,\cdots =1,\cdots,n$. The latter parametrize a $n$-dimensional space that
can be seen as a codimension-1 hypersurface in a $(n+1)$-dimensional space
with coordinates $h^{I}$ and Riemannian metric

\begin{equation}
\label{eq:h_I}
a_{IJ}
=
-2C_{IJK}h^{K} +3h_{I}h_{J}\, ,    
\,\,\,\,\,\,
\mbox{where}
\,\,\,\,\,\,
h_{I}\equiv C_{IJK}h^{J}h^{K}\, ,
\,\,\,\,\,
\Rightarrow
\,\,\,\,\,
h^{I}h_{I}=1\, .
\end{equation}

\noindent
The codimension-1 hypersurface is defined by the cubic equation

\begin{equation}
\label{eq:sp_geom_const}
C_{IJK}h^{I}h^{J}h^{K}=1\, ,  
\end{equation}

\noindent
which will be solved by some parametrization in terms of the physical scalars
$h^{I}(\phi)$. The metric induced in this hypersurface (up to a
normalization factor) is the $\sigma$-model metric for the physical scalars

\begin{equation}
g_{xy}
\equiv
3a_{IJ}\frac{\partial h^{I}}{\partial\phi^{x}} \frac{\partial
  h^{I}}{\partial\phi^{y}}\, .
\end{equation}

It is customary to define 

\begin{equation}
h^{I}_{x} 
\equiv
-\sqrt{3} h^{I}{}_{,x}
\equiv  
-\sqrt{3} \frac{\partial h^{I}}{\partial\phi^{x}}\, ,  
\hspace{1cm}
h_{Ix}
\equiv  
+\sqrt{3}h_{I, x}\, ,
\,\,\,\,\,
\Rightarrow
\,\,\,\,\,
h_{I}h^{I}_{x}
=
h^{I}h_{Ix}
=
0\, ,   
\end{equation}

\noindent
which satisfy\footnote{These two properties can be seen as the definition of
  the metric $a_{IJ}$.}

\begin{equation}
h_{I}  = a_{IJ}h^{I}\, ,
\hspace{1cm}
h_{Ix}  = a_{IJ}h^{J}{}_{x}\, ,
\end{equation}

\noindent
and the completeness relation

\begin{equation}
\label{eq:completeness}
a_{IJ} 
=
h_{I}h_{J}+g_{xy}h^{x}_{I}h^{y}_{J}\, .  
\end{equation}

The geometry defined by these objects is known as \textit{Real Special
  Geometry}.

There are two kinds of global symmetries in these theories: the isometries of
the Real Special manifold and R-symmetry group, which is SU$(2)$. In absence
of hypermultiplets, they can be considered (but not gauged!) independently.
The necessary and sufficient conditions for the gauging of a subgroup of the
global isometry group are:

\begin{enumerate}
\item The subgroup of the isometry group must act on the vector fields
  $A^{I}{}_{\mu}$ in the adjoint representation. This means that we can use
  the same indices $I,J,\ldots$ for the vector multiplets and for the gauge
  group's generators, some of which could be trivial because the isometry
  group does not need to act on all the vector fields. It also means that
  these isometries must act linearly on the functions $h^{I}(\phi)$.

\item It must be a symmetry of the $C_{IJK}$ tensor that defines the
  theory. This condition can be expressed in the form

\begin{equation}
\label{eq:CIJKinvariance}
-3f_{I(J}{}^{M}C_{KL)M}=0\, ,
\end{equation}

\noindent
where $f_{IJ}{}^{K}$ are the gauge group's structure constants,\footnote{These
  structure constants will be trivial in the direction in which the subgroup
  to be gauged does not act.} and it automatically implies the invariance of
the Riemannian metric $a_{IJ}$ under the linear transformations.

\item The functions $h^{I}(\phi)$ must be invariant under those linear
  transformations up to a reparametrization (a field redefinition of the
  scalars). Combined with the above condition, it implies that these
  reparametrizations are isometries of the induced metric $g_{xy}(\phi)$ and
  the vectors that generate them are Killing vectors and must necessarily be
  of the form

\begin{equation}
\label{eq:Killing}
k_{I}{}^{x}= -\sqrt{3}f_{IJ}{}^{K}h_{K}^{x}h^{J}\, .
\end{equation}

This condition eliminates the possiblity of gauging Abelian subgroups of the
isometry group and it is the reason why Abelian gauging is a synonym of
gauging via Abelian Fayet-Iliopoulos terms in these theories. One can
immediately check using the properties of Real Special geometry that these
vectors satisfy the Lie algebra

\begin{equation}
[k_{I},k_{J}]= -f_{IJ}{}^{K}k_{K}\, .
\end{equation}

\end{enumerate}

This kind of symmetries can be gauged immediately by the standard procedure,
giving rise to what have been called $\mathcal{N}=1,d=5$
Super-Einstein-Yang-Mills (SEYM) models, which are the simplest
$\mathcal{N}=1$ supersymmetrization of the $d=5$ Einstein-Yang-Mills system
\cite{Meessen:2015enl}. 

An important property of these theories is that their scalar potential
vanishes identically. Thus, they cannot be used as supersymmetrizations of
EYM-AdS theories. For this purpose one must gauge (a subgroup of) R-symmetry.

R-symmetry (SU$(2)$ in these theories) is always present in any
$\mathcal{N}=1,d=5$ supergravity theory and only acts on the fermions. In
order to gauge the full R-symmetry group, though, we need a triplet of vector
fields transforming in the adjoint of SU$(2)$ and their transformation under
this SU$(2)$ must also be a symmetry of the theory. Since vectors come in
vector multiplets, it is clear that there must be an SU$(2)$ subgroup of the
isometry group that satisfies the above criteria for
\textit{gaugeability}. Gauging the full R-symmetry group, then, involves a
deformation of a SEYM model in which new couplings to the fermions are
introduced in the action, as well as fermion shifts in the supersymmetry
transformation rules and a non-vanishing scalar potential (see
Eq.~(\ref{eq:scalarpotential}) below). Only the latter occurs in the bosonic
action. These new couplings are determined by an object,
$\mathsf{P}_{I}{}^{r}$, ($r,s,...=1,2,3$ are $\mathfrak{su}(2)$ indices) with
only three of the $I$ components non-vanishing\footnote{There is always a
  basis in which this is true.}  satisfying, for some constant $\xi$, the
property\footnote{Here the only non-vanishing components of the structure
  constants $f_{IJ}{}^{K}$ are those of the R-symmetry group SU$(2)$.}

\begin{equation}
\mathsf{P}_{I}{}^{r} = \xi e_{I}{}^{r}\, ,
\hspace{1cm}
\varepsilon^{rst} e_{I}{}^{s} e_{J}{}^{t} 
= 
f_{IJ}{}^{K} e_{K}{}^{r}\, .  
\end{equation}

\noindent
This object plays the r\^ole of an embedding tensor, selecting the three gauge
vectors among the set of all vectors of the theory. It can also be seen as a
constant triholomorphic momentum map. 

The theories obtained by gauging the whole SU$(2)$ R-symmetry group can be
seen as the supersymmetrizations of SU$(2)$-EYM-AdS theories, but, how about
other gauge groups? The only possibility would be to combine a
Fayet-Iliopoulos gauging with the gauging of the desired subgroup of the
global isometry group G of a theory. The resulting theory would have the gauge
group SU$(2)\times$G, but there is a simpler possibility: combining the
gauging of the desired subgroup of the global isometry group G of a theory
with the gauging of a U$(1)$ subgroup of the R-symmetry group using
Fayet-Iliopoulos terms. Gauging a U$(1)$ subgroup of the R-symmetry group is
much simpler, since any vector of the theory can be used as gauge vector. It
will be associated to a $\mathsf{P}_{I}{}^{r}$ with only one $I$-component
different from zero. The resulting theory would have the gauge group
U$(1)\times$G and a scalar potential that, potentially, can give rise an AdS
cosmological constant. This is the kind of gauging that we are going to study
in this paper.\footnote{Supersymmetric solutions of theories in which the
  whole R-symmetry group has been gauged have been studied in
  Ref.~\cite{Ortin:2016sih}.}

It goes without saying that, being completely independent, each of the factors
of the gauge group has its own coupling constant, which we will denote by $g$
for the non-Abelian factor and $g_{0}$ for the Abelian one. The latter will
not appear explicitly in the action that we are about to write because we have
absorbed it into the $\mathsf{P}_{I}{}^{r}$. This is very convenient in the
case we have at hands.

The bosonic action of a theory of $\mathcal{N}=1,d=5$ supergravity coupled to
vector multiplets with the two kinds of gaugings that we have discussed above
is given by

\begin{equation}
\label{eq:generaln1d5action}
\begin{array}{rcl}
S 
& = &  
{\displaystyle\int} d^{5}x\sqrt{g}\
\biggl\{
R
+{\textstyle\frac{1}{2}}g_{xy}\mathfrak{D}_{\mu}\phi^{x}
\mathfrak{D}^{\mu}\phi^{y}
-V(\phi)
-{\textstyle\frac{1}{4}} a_{IJ} F^{I\, \mu\nu}F^{J}{}_{\mu\nu}
\\ \\ & & 
+{\textstyle\frac{1}{12\sqrt{3}}}C_{IJK}
{\displaystyle\frac{\varepsilon^{\mu\nu\rho\sigma\alpha}}{\sqrt{g}}}
\left[
F^{I}{}_{\mu\nu}F^{J}{}_{\rho\sigma}A^{K}{}_{\alpha}
-{\textstyle\frac{1}{2}}gf_{LM}{}^{I} F^{J}{}_{\mu\nu} 
A^{K}{}_{\rho} A^{L}{}_{\sigma} A^{M}{}_{\alpha}
\right.
\\ \\ & & 
\left.
+{\textstyle\frac{1}{10}} g^2 f_{LM}{}^{I} f_{NP}{}^{J} 
A^{K}{}_{\mu} A^{L}{}_{\nu} A^{M}{}_{\rho} A^{N}{}_{\sigma} A^{P}{}_{\alpha}
\right]
\biggr\}\, ,
\end{array}
\end{equation}

\noindent
where $V(\phi)$, the scalar potential, is given by

\begin{equation}
\label{eq:scalarpotential}
V(\phi) 
=  
-\left(4h^{I}h^{J} -2g^{xy}h_{x}^{I}h_{y}^{J}\right)
\mathsf{P}_{I}{}^{r}\mathsf{P}_{J}{}^{r}\, ,
\end{equation}

\noindent
$\mathfrak{D}_{\mu} \phi^{x}$ are the gauge-covariant derivatives of the
scalars

\begin{equation}
\mathfrak{D}_{\mu} \phi^{x} 
= 
\partial_{\mu} \phi^{x}+gA^{I}{}_{\mu} k_{I}{}^{x}\, ,
\end{equation}

\noindent
and $F^{I}{}_{\mu\nu}$ are the gauge-covariant vector field strengths

\begin{equation}
F^{I}{}_{\mu\nu}=2\partial_{[\mu}A^{I}{}_{\nu]}  
+gf_{JK}{}^{I}A^{J}{}_{\mu}A^{K}{}_{\nu}\, .
\end{equation}

The equations of motion are

\begin{eqnarray}
G_{\mu\nu}
-{\textstyle\frac{1}{2}}a_{IJ}\left(F^{I}{}_{\mu}{}^{\rho} F^{J}{}_{\nu\rho}
-{\textstyle\frac{1}{4}}g_{\mu\nu}F^{I\, \rho\sigma}F^{J}{}_{\rho\sigma}
\right)      
& & \nonumber\\
& & \nonumber  \\
\label{eq:eom1}
+{\textstyle\frac{1}{2}}g_{xy}\left(\mathfrak{D}_{\mu}\phi^{x} 
\mathfrak{D}_{\nu}\phi^{y}
-{\textstyle\frac{1}{2}}g_{\mu\nu}
\mathfrak{D}_\rho\phi^{x} \mathfrak{D}^{\rho}\phi^{y}\right)
+{\textstyle\frac{1}{2}}g_{\mu\nu}V
& = &
0\, ,
\\
& & \nonumber \\
\label{eq:eom2}
\mathfrak{D}_{\mu}\mathfrak{D}^{\mu}\phi^{x} 
+{\textstyle\frac{1}{4}}g^{xy} \partial_{y}
a_{IJ} F^{I\, \rho\sigma}F^{J}{}_{\rho\sigma}
+g^{xy}\partial_{y}V
& = &
0\, ,
\\ 
& & \nonumber \\
\label{eq:eom3}
\mathfrak{D}_{\nu}\left( a_{IJ}F^{J\, \nu\mu}\right)
+{\textstyle\frac{1}{4\sqrt{3}}} 
\frac{\varepsilon^{\mu\nu\rho\sigma\alpha}}{\sqrt{g}}
C_{IJK} F^{J}{}_{\nu\rho}F^{K}{}_{\sigma\alpha}
+g
k_{I\,x} \mathfrak{D}^{\mu}\phi^{x}
& = & 
0\, .
\end{eqnarray}

In what remains of this section we are going to describe the two models that
we are going to work with and their gaugings.

\subsection{A simple model with $ \mathrm{SU}(2)\times\mathrm{U}(1)_{\rm R}$
  gauge symmetry}
\label{sec-simplemodel}

As a warm-up exercise one can consider the simplest model that admits a
gauging of the kind we want to consider. It contains a triplet of vector
multiplets labeled by $x,y,z=1,2,3$ and it is defined by the $C_{IJK}$ tensor
with components

\begin{equation}
C_{000}=1\, ,
\quad 
C_{0xy}=-\tfrac{1}{2} \delta_{xy}\, .
\end{equation}

The tensor $C_{IJK}$ tensor\footnote{And, as a consequence, the whole Real
  Special structure. For example, using
\begin{equation}
\label{eq:h_Isimplest}
(h^{0})^{3} -\tfrac{3}{2}h^{0} h^{x}h^{x} = 1\, ,
\hspace{.6cm}
h_{0} = (h^{0})^{2} -\tfrac{1}{2}h^{x}h^{x} = \tfrac{2}{3}(h^{0})^{2}
+\frac{1}{3h^{0}}\, ,
\hspace{.6cm}
h_{x} = - h^{0}h^{x}\, . 
\end{equation}
the components of the kinetic matrix for the vector fields are given by 
\begin{equation}
\label{eq:aIJsimplemodel}
a_{00}
=
\tfrac{4}{3}(h^{0})^{4}  -\tfrac{2}{3} h^{0} + \frac{1}{3(h^{0})^{2}}\, ,
\hspace{.6cm}
a_{0x} 
=
h^{x}[1-2(h^{0})^{3}]\, ,
\hspace{.6cm}
a_{xy} 
=
h^{0}\delta_{xy}+3(h^{0})^{2}h^{x}h^{y}\, .
\end{equation}
Using the coordinates
\begin{equation}
\phi^{x}\equiv \sqrt{\tfrac{3}{2}}h^{x}/h^{0}\, , 
\,\,\,\,\,
\Rightarrow
\,\,\,\,\,
h^{0} = (1-\phi^{2})^{-1/3}\, ,
\,\,\,\,
\mbox{where}
\,\,\,\,
\phi^{2}\equiv \phi^{x} \phi^{x}\, ,  
\end{equation}
these take the form
\begin{equation}
a_{00}
=
\tfrac{4}{3}(h^{0})^{4}  -\tfrac{2}{3} h^{0} + \frac{1}{3(h^{0})^{2}}\, ,
\hspace{.4cm}
a_{0x} 
=
\sqrt{\tfrac{2}{3}}
\phi^{x} h^{0}[1-2(h^{0})^{3}]\, ,
\hspace{.4cm}
a_{xy} 
=
h^{0}\delta_{xy}+2(h^{0})^{4}\phi^{x}\phi^{y}\, ,
\end{equation}
and the $\sigma$-model metric is given by 
%
%
\begin{equation}
g_{xy}
=
\frac{2}{1-\phi^{2}}
\left[\delta_{xy} +\frac{8(3-2\phi^{2})}{9(1-\phi^{2})}\phi^{x}\phi^{y}
\right]\, .  
\end{equation}
} is obviously invariant under SU$(2)$ rotations which act in the adjoint
representation on the triplet of vector multiplets. Therefore, this group of
symemtries can be gauged using the matetr vectors fields $A^{x}{}_{\mu}$ as
gauge fields. The remaining vector field, the graviphoton $A^{0}{}_{\mu}$ can
be used to gauge U$(1)_{\rm R}\subset\mathrm{SU}(2)_{\rm R}$, which, as we
have said, is always possible. More explicitly, we choose

\begin{equation}
\label{eq:momentummapchoice}
\mathsf{P}_{I}{}^{r} 
\equiv 
g_{0} \delta_{I}{}^{0} \delta^{r}{}_{1}\, ,
\end{equation}

\noindent
which includes a choice of the particular specific U$_{\rm
  R}(1)\subset\mathrm{SU}(2)_{\rm R}$ to be gauged.

The only manifestation of this gauging in the bosonic action
Eq.~(\ref{eq:generaln1d5action}) is the presence of the scalar potential,
whose explicit form we will not be concerned with. Furthermore,

\begin{equation}
F^{0}{}_{\mu\nu}= 2\partial_{[\mu}A^{0}{}_{\nu]}\, .  
\end{equation}

\noindent
The covariant derivatives of the scalars and the vector field strengths refer
to the SU$(2)$ gauging and are explicitly given by\footnote{The structure
  constants and Killing vectors are given by
\begin{equation}
f_{xy}{}^{z}=\varepsilon_{xyz}\, , 
\hspace{1cm}
k_{x}{}^{y} =\varepsilon_{xyz}\phi^{z}\, .   
\end{equation}
} 

\begin{equation}
  \begin{array}{rcl}
\mathfrak{D}_{\mu} \phi^{x} 
& = &
\partial_{\mu} \phi^{x}+g \varepsilon^{xyz}A^{z}{}_{\mu} \phi^{z}\, ,
\\
& & \\
F^{x}{}_{\mu\nu}
& = &
2\partial_{[\mu}A^{x}{}_{\nu]}  
+g\varepsilon^{xyz}A^{z}{}_{\mu}A^{y}{}_{\nu}\, .
\end{array}
\end{equation}

\subsection{The $\mathbb{C}$ magic model with $\mathrm{SU}(3)\times$U$(1)_{\rm
    R}$ gauge symmetry}
\label{sec-themodel}

The second model that we are going to consider is the so-called ``$\mathbb{C}$
magic model'', associated with the ``magic'' Jordan algebra
$\mathbf{J}_{3}^{\mathbb{C}}$ \cite{Gunaydin:1983bi}. This model is one of the
possible truncations of maximal $d=5$ supergravity and is one of the symmetric
Real Special geometries \cite{Gunaydin:1983rk}. Furthermore, in
Ref.~\cite{Gunaydin:1985cu} it was shown that the maximal $d=5$ supergravity
with SO(6) gauging can be consistenly truncated to this model with an
SU$(3)\times\mathrm{U}(1)_{\rm R}$ gauging (a model previously constructed in
Ref.~\cite{Gunaydin:1984ak}), which belongs to the class we want to consider
in this paper.

The $\mathbb{C}$ magic model  is determined by the constant symmetric tensor
$C_{IJK}$ of non-vanishing components

\begin{equation}
C_{000}=1\, ,
\quad 
C_{0xy}=-\tfrac{1}{2} \delta_{xy}\, ,
\quad 
C_{xyz}= \sqrt{\tfrac{3}{8}}\, d_{xyz}\,,
\end{equation}

\noindent
where $x,y,z=1,\ldots,8$ and $d_{xyz}$ is the fully symmetric constant tensor
associated with SU$(3)$, given in terms of the Gell-Mann matrices
$\lambda_{x}$ as

\begin{equation}
 d_{xyz} = \tfrac{1}{2} \Tr\left[\lambda_{x}\{\lambda_{y},\lambda_{z}\}\right]\,,
\end{equation}

\noindent
and having non-vanishing components

\begin{equation}
\begin{array}{c}
d_{146}=d_{157}=d_{256}=d_{344}=d_{355}=1\, ,
\quad 
d_{247}=d_{366}=d_{377}=-1
\\
\\
d_{118}=d_{228}=d_{338}=\tfrac{2}{\sqrt{3}}\, ,
\quad 
d_{448}=d_{558}=d_{668}=d_{778}=-\tfrac{1}{\sqrt{3}}\, ,
\quad 
d_{888}=-\frac{2}{\sqrt{3}}\, .
\end{array}
\end{equation}

It can be seen that the scalar fields parametrize the symmetric space
SL$(3,\mathbb{C})$/SU$(3)$.  The gauge fields $A^{x}$ transform in the adjoint
representation of SU($3$), the maximal compact subgroup of the scalar group
manifold, as do the scalar functions $h^{x}$ and, therefore, they can be used
as SU$(3)$ gauge fields.  $A^{0}$ gauges U$(1)_{\rm R}\subset\mathrm{SU}_{\rm
  R}(2)$.  Without any loss of generality, we select this subgroup as in
Eq.~(\ref{eq:momentummapchoice}).

Observe that, being a symmetric model, with the normalization chosen here,

\begin{equation}
\label{eq:symmprop}
C^{IJK}=C_{IJK}\, .  
\end{equation}

We will be interested in solution in which a only a subgroup SU$(2)\subset
\mathrm{SU}(3)$ is active. However, it turns out that an additional U$(1)$
must also remain active. 

\section{Timelike supersymmetric solutions}
\label{sec-susysolutions}

The supersymmetric solutions of matter-coupled $\mathcal{N}=1,d=5$
supergravity theories with arbitrary gaugings have been fully characterized in
a series of papers in which couplings of increasing complexity were considered
\cite{Gauntlett:2002nw,Gauntlett:2003fk,Gutowski:2004yv,Gauntlett:2004qy,Gutowski:2005id,Bellorin:2006yr,Bellorin:2007yp,Bellorin:2008we}. Using
these characterizations one can define procedures to construct, step by step,
supersymmetric solutions.  These procedures have become extremely useful
solution-generating techiques.

We are going to search for timelike supersymmetric solutions of the two
models reviewed in Sections~\ref{sec-simplemodel} and
\ref{sec-themodel}.  For this case it turns out that we can simply reuse the
procedure described in Ref.~\cite{Chimento:2017sen} for Abelian gaugings,
coveniently covariantized to include the non-Abelian gauging. The
solution-generating recipe is in full agreement with the general recipe
obtained in the above-mentioned references and, before we specify the choice
of momentum maps, it can be summarized as follows:

First of all, the building blocks of the  timelike supersymmetric
solutions are

\begin{enumerate}
\item The 4-dimensional spatial metric $h_{\underline{m}\underline{n}}$, where
  $\underline{m},\underline{n},\underline{p}=1,\cdots,4$.\footnote{In our
    conventions, underlined indices are world indices. Tangent-space indices
    will not be underlined.} It does not depend on the time coordinate and
  defines a 4-dimensional spatial manifold usually called ``base space'' which
  plays an auxiliary r\^ole and has no direct physical relevance.  All the
  building blocks and operators used in what follows are naturally defined in
  this 4-dimensional space and, hence, they are time-independent. We use hats
  to denote them.

\item The antiselfdual almost hypercomplex structure
  $\hat{\Phi}^{(r)}{}_{mn}$, $r,s,t=1,2,3$. By definition, the 2-forms
  satisfy the properties

  \begin{eqnarray}
    \hat{\Phi}^{(r)\, mn} 
    & = &
    -\tfrac{1}{2}\varepsilon^{mnpq}\hat{\Phi}^{(r)}{}_{pq}\, , 
    \hspace{1cm}
    \mbox{or}
    \hspace{1cm}
    \hat{\Phi}^{(r)}=-\hat{\star}\hat{\Phi}^{(r)}\, ,
    \\
    & & \nonumber \\
    \hat{\Phi}^{(r)\, m}{}_{n}  \hat{\Phi}^{(s)\, n}{}_{p}
    & = &
    -\delta^{rs} \delta^{m}{}_{p} 
    +\varepsilon^{rst}  \hat{\Phi}^{(t)\, m}{}_{p}\, .
  \end{eqnarray}

\item The scalar function $\hat{f}$.

\item The 1-form $\hat{\omega}_{\underline{m}}$.

\item The 1-form potentials $\hat{A}^{I}{}_{\underline{m}}$.

\item The functions of the physical scalars $h^{I}(\phi)$. They are
  time-independent as well.

\end{enumerate}

These building blocks must fulfill the following conditions:

\begin{enumerate}

\item The antiselfdual almost hypercomplex structure
  $\hat{\Phi}^{(r)}{}_{mn}$, the 1-form potentials
  $\hat{A}^{I}{}_{\underline{m}}$ and the base-space metric
  $h_{\underline{m}\underline{n}}$ (through its Levi-Civita connection) must
  solve the equation\footnote{The local SU$(2)$ symmetry of this differential
    equation is formally that of the full SU$_{\rm R}(2)$ until the values of the
    momentum maps $\mathsf{P}_{I}{}^{s}$ are specified. After the choice
    Eq.~(\ref{eq:momentummapchoice}) this differential equation splits into
    Eqs.~(\ref{eq:df1=0})-(\ref{eq:df3=-pf2}). We are going to discuss the
    specifics of the models we are considering next.}

\begin{equation}
\label{eq:dfAf}
\hat{\nabla}_{m}\hat{\Phi}^{(r)}{}_{np} 
+
\varepsilon^{rst}\hat{A}^{I}{}_{m}\mathsf{P}_{I}{}^{s}\hat{\Phi}^{(t)}{}_{np}
=0\, .
\end{equation}

\item The selfdual part of the spatial vector field strengths
  $\hat{F}^{I}\equiv d\hat{A}^{I}+\tfrac{1}{2}gf_{JK}{}^{I}\hat{A}^{J}\wedge
  \hat{A}^{K} $ is given by

\begin{equation}
\label{eq:susy_{2}}
h_{I}\hat{F}^{I+} 
= 
{\textstyle\frac{2}{\sqrt{3}}} (\hat{f}d\hat{\omega})^{+} \, . 
\end{equation}

\item The antiselfdual part of $\hat{F}^{I}$ is given by\footnote{In this
    equation the indices of $C^{IJK}$ have been raised using the inverse
    metric $a^{IJ}$. This object satisfies the relations
\begin{equation}
\label{eq:Ch}
C^{IJK}h_{K}
=
h^{I}h^{J} -\tfrac{1}{2}g^{xy}h_{x}^{I}h_{y}^{J}
=\tfrac{3}{2}h^{I}h^{J}-\tfrac{1}{2}a^{IJ}\, ,
\end{equation}
the first of which allow us to rewrite the scalar potential in
Eq.~(\ref{eq:scalarpotential}) in the form 
\begin{equation}
\label{eq:scalarpotentialalt}
V(\phi) 
=  
-4
C^{KIJ}h_{K}\mathsf{P}_{I}{}^{r}\mathsf{P}_{J}{}^{r}\, .
\end{equation}
}

\begin{equation}
\label{eq:Fminus}
\hat{F}^{I-} 
=
-2\hat{f}^{-1}C^{IJK}h_{J}\mathsf{P}_{K}{}^{r}\hat{\Phi}^{(r)}\, .
\end{equation}

\item Finally, the following equation relating all the building blocks, where
  the dots indicate standard contraction of all the indices of the tensors,
  has to be satisfied

\begin{equation}
\label{eq:susy_4}
\hat{\mathfrak{D}}^{2}\left(h_{I}/\hat{f}\right)
-\tfrac{1}{6}C_{IJK}
\hat{F}^{J}\cdot\hat{\star}\hat{F}^{K}
+{\textstyle\frac{1}{2\sqrt{3}}}
\left(a_{IK}-2C_{IJK}h^{J}\right)\hat{F}^{K}\cdot (\hat{f}d\hat{\omega})^{-}
=
0\, .     
\end{equation}

\end{enumerate}

Having found building blocks that satisfy the above conditions, the physical
5-dimensional fields are reconstructed as follows:

\begin{enumerate}
\item The 5-dimensional metric is given by

\begin{equation}
\label{eq:metric_form}
  ds^{2} 
  = 
  \hat{f}^{\, 2}(dt+\hat{\omega})^{2}
  -\hat{f}^{\, -1}h_{\underline{m}\underline{n}}dx^{m} dx^{n}\, .
\end{equation}

\item The complete 5-dimensional vector fields are given by

\begin{equation}
  \label{eq:completevectorfields}
  A^{I} 
  = 
  -\sqrt{3}h^{I}e^{0} +\hat{A}^{I}\, ,    
  \,\,\,\,\,
  \mbox{where}
  \,\,\,\,\,
  e^{0} 
  \equiv
  \hat{f} (dt +\hat{\omega})\, .
\end{equation}



\noindent
The complete 5-dimensional field strength is given by 

\begin{equation}
\label{eq:gauge_field_strenghts}
F^{I}  
= 
-\sqrt{3} \hat{\mathfrak{D}}(h^{I} e^{0})  +\hat{F}^{I}\, .
\end{equation}

\item The scalar fields $\phi^{x}$ can be obtained by inverting the functions
  $h_{I}(\phi)$ or $h^{I}(\phi)$ is the form of these functions is known. One
  can always use a parametrization of these functions such that the scalars
  are given by  

\begin{equation}
\label{eq:phys_scalars}
  \phi^{x}= h_{x}/h_{0}=(h_{x}/\hat{f})/(h_{0}/\hat{f})\, .    
\end{equation}

\end{enumerate}

When we specify the U$(1)_{\rm R}\subset\mathrm{SU}(2)_{\rm R}$ that we are
going to gauge and corresponding gauge vector as in
Eq.~(\ref{eq:momentummapchoice}) it is possible to extract more information
from the equations satisfied by the building blocks of timelike supersymmetric
solutions. We analyze them next.

\subsection{Supersymmetric solutions of cosmological gauged models}
\label{sec-susysolutionscomogauged}

With the choice Eq.~(\ref{eq:momentummapchoice}), Eq.~(\ref{eq:dfAf}) splits
into the following three equations \cite{Gauntlett:2003fk,Gutowski:2005id}

\begin{eqnarray}
\label{eq:df1=0}
\hat{\nabla}_{m}\hat{\Phi}^{(1)}{}_{np} 
& = &
0\, , 
\\
& & \nonumber \\
\label{eq:df2=pf3}
\hat{\nabla}_{m}\hat{\Phi}^{(2)}{}_{np} 
& = & 
g_{0}\hat{A}^{0}{}_{m}\hat{\Phi}^{(3)}{}_{np}\, ,
\\
& & \nonumber \\
\label{eq:df3=-pf2}
\hat{\nabla}_{m}\hat{\Phi}^{(3)}{}_{np} 
& = & 
-g_{0}\hat{A}^{0}{}_{m}\hat{\Phi}^{(2)}{}_{np}\, ,
\end{eqnarray}

\noindent
the first of which implies that the ``base space'' metric
$h_{\underline{m}\underline{n}}$ is K\"ahler with respect to the complex
structure $\hat{J}_{mn}\equiv\hat{\Phi}^{(1)}{}_{mn}$. Then, the integrability
condition of the other two equations leads to a relation between the U$_{\rm
  R}(1)$ gauge potential and the base space metric

\begin{equation}
\label{eq:R=dP}
\hat{\mathfrak{R}}_{mn} 
=
-g_{0}\hat{F}^{0}{}_{mn}\, ,  
\end{equation}

\noindent
where $\hat{\mathfrak{R}}_{mn}$ is the Ricci 2-form of the K\"ahler base
space.

Eq.~(\ref{eq:Ch}) is not simplified by our choice of gauging, but
Eq.~(\ref{eq:Fminus}) becomes

\begin{equation}
\label{eq:susy_{3}}
\hat{F}^{I-}
=
-2g_{0}\hat{f}^{-1}C^{0IJ}h_{J}\hat{J}\, ,
\end{equation}

\noindent
which implies 

\begin{eqnarray}
\label{eq:F0-}
\hat{F}^{0-} 
& = & 
{\displaystyle\frac{1}{2 g_{0}}}\hat{f}^{-1}V(\phi) \hat{J}\, ,
\\
& & \nonumber \\
h_{I}\hat{F}^{I-}
& = &
-2g_{0}\hat{f}^{-1}h^{0}\hat{J}\, .
\end{eqnarray}

\noindent
Then, the trace of Eq.~(\ref{eq:R=dP}) and Eq.~(\ref{eq:F0-}) with
$\hat{J}^{mn}$ together lead to

\begin{equation}
\label{eq:RicciVf}
\hat{R}= - 2V(\phi)/\hat{f}\, .
\end{equation}

Finally, substituting Eq.~(\ref{eq:susy_{3}}) into Eq.~(\ref{eq:susy_4}) and
using in it Eqs.~(\ref{eq:Ch}) and (\ref{eq:completeness}), and taking into
account that $h_{0}$ is a singlet under the non-Abelian factor of the gauge
group, we get the following two equations

\begin{eqnarray}
\label{eq:susy_4-2-1}
\hat{\nabla}^{2}(h_{0}/\hat{f})
-\tfrac{1}{6}C_{0JK}
\hat{F}^{J}\cdot[\hat{\star}\hat{F}^{K} +4\sqrt{3}h^{K} (\hat{f}d\hat{\omega})^{-}]
-\sqrt{3}g_{0}h_{0}h^{0}\hat{J}\cdot d\hat{\omega}
& = &
0\, ,
\\
& & \nonumber \\ 
\label{eq:susy_4-2-2}
\hat{\mathfrak{D}}^{2}(h_{x}/\hat{f})
-\tfrac{1}{6}C_{xJK}
\hat{F}^{J}\cdot[\hat{\star}\hat{F}^{K} +4\sqrt{3}h^{K} (\hat{f}d\hat{\omega})^{-}]
-\sqrt{3}g_{0}h_{x}h^{0}\hat{J}\cdot d\hat{\omega}
& = &
0\, . 
\end{eqnarray}

In order to simplify the construction of solutions of this class, which should
start by judicious choice of the 4-dimensional K\"ahler metric, we are going to
assume that this K\"ahler metric admits a holomorphic isometry. Then, it can
always be written as\footnote{See Ref.~\cite{Chimento:2016run} and references
  therein.}

\begin{equation}
\label{eq:Kahlerisometric} 
d\hat{s}_{4}^{2}
=
h_{\underline{m}\underline{n}}dx^{m}dx^{n} 
= 
H^{-1}\left( dz+\chi \right)^{2}
+H\left\{(dx^{2})^{2}+W^{2}(\vec{x})[(dx^{1})^{2}+(dx^{3})^{2}]\right\}\, ,
\end{equation} 

\noindent
with the functions $H$ and $W$, and the 1-form $\chi$, independent of the
coordinate $z$, which is adapted to the holomorphic isometry, and satisfying
the constraint

\begin{equation}
\label{eq:constraintijcurved}
\breve{\star}_{3} d\chi =dH+H\partial_{\underline{2}} \log{W^{2}} dx^{2}\, ,
\end{equation}

\noindent
where $\breve{\star}_{3}$ is the Hodge dual in the 3-dimensional manifold 

\begin{equation}
\label{eq:tridimensionalmetric}
d\breve{s}^{2}_{3}
=
(dx^{2})^{2}+W^{2}(\vec{x})[(dx^{1})^{2}+(dx^{3})^{2}]\, .
\end{equation}

\noindent
The integrability condition of the constraint
Eq.~(\ref{eq:constraintijcurved}) is the equation

\begin{equation}
\label{eq:integrabilityconditon}
\partial_{\underline{1}}\partial_{\underline{1}} H
+\partial_{\underline{2}}\partial_{\underline{2}}(W^{2} H)
+\partial_{\underline{3}}\partial_{\underline{3}} H 
= 0\, .
\end{equation}

Therefore, the simplifying assumption of the existence of a holomorphic
isometry allows us to construct any K\"ahler metric within this class by
choosing an arbitrary function $W$, solving the integrability condition
Eq.~(\ref{eq:integrabilityconditon}) for $H$ and then solving the constraint
Eq.~(\ref{eq:constraintijcurved}) for $\chi$.

In order to make progress it is necessary to specify the model under
consideration. We start by the simple model described in
Section~\ref{sec-simplemodel}.

\subsection{Supersymmetric solutions of the simplest $\mathrm{SU}(2)\times
  \mathrm{U}(1)_{\rm R}$ model}
\label{sec-susysolutionssimplemodel}

Here we are going to label the three vector multiplets with $
A,B,\ldots=1,\ldots,3$ to simplify the comparison with the $\mathbb{C}$ magic
model, which will have an SU$(2)$ triplet of vectors active but has more
vector multiplets labeled, according with the general notation, by
$x,y\ldots=1,\ldots, n_{V}$.

For the sake of simplicity, we are going to impose

\begin{equation}
\hat{F}^{A-}=0\, ,\quad \text{and} \quad h_{A}=0\, .
\end{equation}

\noindent
Then, the SU$(2)$ gauge field is a selfdual instanton on the K\"ahler base
space and we can use the results of Ref.~\cite{Chimento:2017tsv} to construct
it.  Also, because of Eqs.~(\ref{eq:h_Isimplest}) we have that $h^{A}=0$ and
$h_{0}=h^{0}=1$ which, because of Eq.~(\ref{eq:aIJsimplemodel}), imply in
their turn that $a_{IJ}=a^{IJ}=\delta_{IJ}$ and $C^{IJK}=C_{IJK}$.

Then, under these assumptions, Eq.~(\ref{eq:susy_{3}}) takes the form

\begin{equation}
\label{eq:F0-2}
\hat{F}^{0-}
=
-2g_{0}\hat{f}^{-1}\hat{J}\, ,
\end{equation}

\noindent
while Eq.~(\ref{eq:susy_{2}}) gives

\begin{equation}
 \hat{F}^{0+}=\tfrac{2}{\sqrt{3}}( \hat{f} d\hat{\omega} )^{+}\, .
\end{equation}

\noindent
Finally, Eqs.~(\ref{eq:susy_4-2-1}) and (\ref{eq:susy_4-2-2}) take the form 

\begin{eqnarray}
\label{eq:h0simplest}
\hat{\nabla}^{2}\hat{f}^{-1}
-\tfrac{1}{6}
\hat{F}^{0}\cdot\hat{\star}\hat{F}^{0}
+\tfrac{1}{12}
\hat{F}^{A}\cdot\hat{F}^{A}
+\tfrac{1}{\sqrt{3}}g_{0}\hat{J}\cdot d\hat{\omega}
& = &
0\, ,
\\
& & \nonumber \\ 
\label{eq:selfd_dotprod}
\hat{F}^{A}\cdot\hat{\star}\hat{F}^{0}
& = &
0\,
\,\,\,\,
\Rightarrow
\,\,\,\,
\hat{F}^{A+}\cdot (d\hat{\omega})^{+}
=
0\, , 
\end{eqnarray}

\noindent
where we have used the previous equations in both equations.  The simplest way
to solve the last equation is to require\footnote{Actually, following the
  treatment in Ref.~\cite{Chimento:2016mmd} one can show that if one chooses a
  K\"ahler metric admitting a holomorphic isometry, which can be put in the form
  explained in Ref.~\cite{Chimento:2016run} with $H=H(\varrho)$,
  $W^{2}=\Psi(\varrho)\Phi(x^{1},x^{3})$ and $\hat{f}=\hat{f}(\varrho)$, as we
  are going to assume here, then $\hat{F}^{0+}\propto
  \hat{V}^{\sharp}\wedge\hat{V}^{2} +\hat{V}^{3}\wedge\hat{V}^{1}$. It follows
  that for Eq.~(\ref{eq:selfd_dotprod}) to be satisfied, either
  $\hat{F}^{0+}=0$ or $\hat{F}^{A+}_{\sharp 2}=0\ \forall A$.}

\begin{equation}
(d\hat{\omega})^{+}=\hat{F}^{0+}= 0\, .
\end{equation}

Given that $d\hat J=d\hat{F}^{0-}=0$, Eq.~(\ref{eq:F0-2}) implies that $\hat{f}$
is constant, and we can substitute Eq.~(\ref{eq:F0-2}) in
Eq.~(\ref{eq:h0simplest}) obtaining ($\hat{J}\cdot \hat{J}=4$)

\begin{equation}
\label{eq:h0simplest2}
\tfrac{8}{3}g_{0}^{2}\hat{f}^{-2} 
+\tfrac{1}{12}
\hat{F}^{A}\cdot\hat{F}^{A}
+\tfrac{1}{\sqrt{3}}g_{0}\hat{J}\cdot d\hat{\omega}
=
0\, ,  
\end{equation}

\noindent
and also in Eq.~(\ref{eq:R=dP}), which using the results in Appendix~B of
Ref.~\cite{Chimento:2016mmd} leads to the equations

\begin{align}
 \partial_{\underline 1,\underline 3}\left( H^{-1}\partial_\varrho \log W^{2} \right)&=0\,,\\
 \nonumber\\
 \partial_\varrho\left( H^{-1}\partial_\varrho\log W^{2} \right)&=4 g_{0}^{2}\hat{f}^{-1}\,,\\
 \nonumber\\
 \hat\nabla^{2}\log W^{2}&=8 g_{0}^{2}\hat{f}^{-1}\, ,
\end{align}

\noindent
for the functions $H$ and $W$ that appear in the K\"ahler metric Eq.~(\ref{eq:Kahlerisometric}).

The first of these equations is automatically solved if we consider the usual
ansatz $H=H(\varrho)$ ($\varrho\equiv x^{2}$) and
$W^{2}=\Psi(\varrho)\Phi_{(k)}(x^{1},x^{3})$. The integrability condition
Eq.~(\ref{eq:integrabilityconditon}) is then satisfied if

\begin{equation}
\label{eq:HPsi}
H(\varrho)=\frac{\rho^{\epsilon}}{\Psi(\varrho)}\, ,
\,\,\,\,\,
\epsilon=0,1\, ,
\end{equation}

\noindent
where we have used the freedom to shift $\varrho$ and to rescale in opposite way the functions $\Psi$ and $\Phi_{(k)}$. From now on we
consider the $\epsilon=1$ case, which will be the one that will give an
interesting solution (a supersymmetric 1-parameter deformation of AdS$_{5}$).

The remaining equations are solved if $\Psi(\varrho)$ is of the form

\begin{equation}
\label{eq:Psisimplestmodel}
\Psi(\varrho)=\frac{4}{3}g_{0}^{2}\hat{f}^{-1}\varrho^{3}+k\varrho^{2}
+\alpha \ ,
\end{equation}

\noindent
and if $\Phi_{(k)}$ is a solution of Liouville's equation

\begin{equation}
(\partial^{2}_{\underline{1}}+\partial^{2}_{\underline{1}})\log \Phi_{(k)}=-2k\Phi_{(k)}\, ,
\end{equation}

\noindent
with $k$ constant and $\alpha$ is an arbitrary integration constant. 

For $k=+1$ and $\alpha=0$ the base space is the Bergman space
$\overline{\mathbb{CP}}^{2}$.

The only equations left to solve are (\ref{eq:h0simplest2}) plus the
selfduality condition of the non-Abelian field strength $\hat{F}^{A-}=0$.  We
need to solve the latter first, but we make the following observation: if we
find a selfdual SU$(2)$ instanton such that $\hat{F}^{A+}\cdot
\hat{F}^{A+}\equiv 32 g_{0}^{2} \hat{f}^{-2} \lambda$ where $\lambda$ is a
constant, then Eq.~(\ref{eq:h0simplest2}) can be solved by taking
$d\hat{\omega} = -\frac{2}{\sqrt{3}}g_{0}(1+\lambda)\hat{f}^{-2}\hat{J}$,
or, up to a closed form,

\begin{equation}
\hat{\omega} 
= 
\frac{2 g_{0}}{\sqrt{3}}(1+\lambda)\hat{f}^{-2}\varrho(dz+\chi_{(k)})\, .
\end{equation}

If this solution exists, then it is not difficult to see that the full
5-dimensional metric is invariant under the rescaling $t\to t/\sigma$,
$\varrho\to \sigma\varrho$, $\hat{f}\to \sigma \hat{f}$, $\alpha\to \sigma^{2}
\alpha$, which we can use to set $\hat{f}=1$.

Then, we now focus on finding a selfdual SU$(2)$ instanton on the K\"ahler
base spaes that we have just determined through $H$ and $W$ with constant
instanton number density $\hat{F}^{A+}\cdot \hat{F}^{A+}$.

Selfdual SU$(2)$ instantons in 4-dimensional K\"ahler spaces with one
holomorphic isometry have recently been studied in
Ref.~\cite{Chimento:2017tsv}, where a Kronheimer-type relation between those
instantons and monopoles satisfying a generalization of the Bogomol'nyi
equation was found and a subsequent generalization of the \textit{hedgehog
  ansatz} was used to solve the latter in the spherically-symmetric case
$k=+1$.

Let us summarize this result: 

\begin{enumerate}
\item Decomposing the gauge field with respect to the action of the
  holomorphic isometry as

  \begin{equation}
    \label{eq:kro_ansatz}
    \hat{A}^{A}=-H^{-1}\Phi^{A} (dz+\chi_{(1)})+\breve{A}^{A}\, ,
  \end{equation}

\noindent
where $\Phi^{A}$ and $\breve{A}^{A}$ are independent of $z$ and are defined in
the 3-dimensional space with metric Eq.~(\ref{eq:tridimensionalmetric}), and
$H(\varrho)$ is one of the functions that occurs in the generic K\"ahler
metric Eq.~(\ref{eq:Kahlerisometric}) and where it is assumed that $W=
\Psi(\varrho)\Phi_{(1)}$.

\item Assuming in addition that they have the \textit{hedgehog} form

\begin{equation}
  \Phi^{A}
  =
  F(\varrho)\frac{y^{A}}{\varrho}\, , 
  \qquad \text{and}\qquad 
  \breve{A}^{A}
  =
  L(\varrho)\varepsilon^{A}{}_{BC}\frac{y^{B}}{\varrho} 
  d\left(\frac{y^{C}}{\varrho}\right)\, ,
\end{equation}

\noindent
where $y^{A}$ are Cartesian coordinates related to $\varrho$ by
$y^{A}y^{A}=\varrho^{2}$ and $F(\varrho)$ and $L(\varrho)$ are two functions
to be determined,

\end{enumerate}

\noindent
the field strength $F^{A}$ will be selfdual in the 4-dimensional K\"ahler
space with metric Eq.~(\ref{eq:Kahlerisometric}) and $H=H(\varrho),W=
\Psi(\varrho)\Phi_{(1)}$ if the following two equations are satisfied

\begin{equation}
\label{eq:kro_eqs}
\left\{
\begin{array}{rcl}
K' & = & G-1\, ,\\
& & \\
\Psi G' & = & 2 K G\, ,
\end{array}
\right.
\,\,\,\,\,
\text{where}
\,\,\,\,\,
K\equiv g \Psi F\, ,
\,\,\, 
\text{and}
\,\,\,
G\equiv (1+g L)^{2}\, .
\end{equation}

These equations depend explicitly on the function $\Psi(\varrho)$, which in
our case is given by Eq.~(\ref{eq:Psisimplestmodel}) and, precisely for the
$k=+1$ and $\alpha=0$, when the base space is the Bergman space
$\overline{\mathbb{CP}}^{2}$, one of the solutions found in
Ref.~\cite{Chimento:2017tsv} has constant instanton number density, as we were
looking for. This solution is

\begin{equation}
K
=
\tfrac{2}{3}g_{0}^{2}\varrho^{2} \, , 
\hspace{1cm}
G
=
1+\tfrac{4}{3}g_{0}^{2}\varrho\, ,
\end{equation}

\noindent
and its instanton number density is given by 

\begin{equation}
\hat{F}^{A}\cdot \hat{F}^{A}
=
\frac{16}{3}\left(\frac{g_{0}^{2}}{g}\right)^{2}\, ,
\,\,\,\,
 \Rightarrow
\,\,\,\, 
\lambda=\frac{g_{0}^{2}}{6 g^{2}}\, .
\end{equation}

Summarizing: we have found a simple solution whose only non-vanishing fields
are a selfdual $\mathrm{SU}(2)$ instanton living on the base space
$\overline{\mathbb{CP}}^{2}$ plus an Abelian vector field and the metric. The
last two fields take the form

\begin{equation}
\label{eq:quasiads5k1}
\begin{array}{rcl}
ds^{2} 
& = &
\left[ 
dt +\tfrac{2}{\sqrt{3}}g_{0}(1+\lambda) \varrho (dz+\cos{\theta}d\varphi)    
\right]^{2}
\\
& & \\
& & 
-\varrho [1+\tfrac{4}{3}g_{0}^{2}\varrho](dz+\cos{\theta}d\varphi)^{2}
-{\displaystyle\frac{d\varrho^{2}}{\varrho[1+\tfrac{4}{3}g_{0}^{2}\varrho]}}
-\varrho\, d\Omega^{2}_{(2,1)}\, ,
\\
& & \\
F^{0}
& = & 
2\lambda g_{0} \hat{J}\, .
\end{array}
\end{equation}

In the $g\rightarrow \infty$ limit, $\lambda\rightarrow 0$, the Abelian and
non-Abelian gauge fields vanish and the metric is that of $\mathrm{AdS}_{5}$.

\subsection{Supersymmetric solutions of the $\mathrm{SU}(3)\times
  \mathrm{U}(1)_{\rm R}$-gauged $\mathbb{C}$ magic model}
\label{sec-susysolutionsCmagicmodel}

Let us now consider the model presented in Section~\ref{sec-themodel}.  We
start by assuming, for the sake of simplicity,

\begin{equation}
\label{eq:ansatz}
A^{x}{}_{\mu}=0\, ,
\,\,\,\,
\text{and}
\,\,\,\,
h_{x}=0\, ,
\,\,\,\,
3<x<8\, ,
\end{equation}

\noindent
so that we are effectively considering a theory with only four vector
multiplets and gauge group $\mathrm{SU}(2)\times\mathrm{U}(1)_{\rm R}$ with an
extra $\mathrm{U}(1)$ which is ungauged (nothing is charged under it). It
should be stressed that this is not a truncation, but an Ansatz that produces
an important simplification to be tried in the equations. As in the previous
case, we will use indices $A=1,2,3$ for the first three vector multiplets that
gauge the $\mathrm{SU}(2)$ factor. The $\mathrm{U}(1)_{\rm R}$ factor will be
gauged by $A^{0}{}_{\mu}$ and the other surviving vector multiplet corresponds
to $A^{8}{}_{\mu}$.

We are also going to look for solutions containing a selfdual
$\mathrm{SU}(2)$ instanton on the 4-dimensional K\"ahler space and, therefore,
we impose

\begin{equation}
\hat{F}^{A-}=0\, .
\end{equation}

The Ansatz, together with Eq.~(\ref{eq:susy_{3}}) and Eq.~(\ref{eq:symmprop})
implies 

\begin{equation}
h_{A}=0\, , 
\end{equation}

\noindent
which in turn implies that 

\begin{equation}
 h^{x}=0\, ,\quad \forall x=1,\ldots,7\, ,
\end{equation}

\noindent
so that the only non-vanishing scalar functions $h^{I}$ are $h^{0},h^{8}$ and
are related to $h_{0},h_{8}$ by

\begin{equation}
h^{0}=(h_{0})^{2}-\tfrac{1}{2} (h_{8})^{2}\, ,
\quad 
h^{8}=-h_{8}(h_{0}+\tfrac{1}{\sqrt{2}}h_{8})\, .
\end{equation}

\noindent
Furthermore, they satisfy the constraint

\begin{equation}
\label{eq:cubic_constr08}
 (h_{0}-\sqrt{2} h_{8})(h_{0}+\tfrac{1}{\sqrt{2}}h_{8})^{2} 
=(h^{0}-\sqrt{2} h^{8})(h^{0}+\tfrac{1}{\sqrt{2}}h^{8})^{2}=1\, .
\end{equation}

It follows that the non-vanishing components of the metric $a_{IJ}$ are

\begin{equation}
\begin{array}{c}
a_{00}=(h_{0})^{2}+(h_{8})^{2}\, ,
\quad
a_{08}=h_{8}(2h_{0}-\tfrac{1}{\sqrt{2}}h_{8})\, ,
\quad 
a_{88}= (h_{0})^2-\sqrt{2} h_{0} h_{8} + \tfrac{3}{2}(h_{8})^{2}
\\
\\
a_{AB}=\delta_{AB}(h_{0}+\tfrac{1}{\sqrt{2}}h_{8})^{2}\, ,
\hspace{1cm}
\forall A,B=1,2,3\, ,
\\
\\
a_{xy}=\delta_{xy}[(h_{0})^{2}-\tfrac{1}{\sqrt{2}}h_{0}h_{8}-(h_{8})^{2}]\, ,
\hspace{1cm}
\forall x,y=4,\ldots,7\, .
\\
\end{array}
\end{equation}

From the same equations one has

\begin{equation}
\label{eq:F0-F8-}
\hat{F}^{0-}=-2 g_{0} (h_{0}/\hat{f})\, \hat{J}\, ,
\hspace{1.5cm}
\hat{F}^{8-}= g_{0} (h_{8}/\hat{f})\, \hat{J}\,.
\end{equation}

\noindent
while equation (\ref{eq:susy_{2}}) gives

\begin{equation}
\label{eq:hF+08}
h_{0}\hat{F}^{0+}+h_{8} \hat{F}^{8+}
=
\tfrac{2}{\sqrt{3}}( \hat{f} d\hat{\omega} )^{+}\,.
\end{equation}

After using Eqs.~(\ref{eq:cubic_constr08}) and (\ref{eq:F0-F8-}),
Eq.~(\ref{eq:susy_4-2-1}) takes the form

\begin{equation}
\label{eq:Max0}
\begin{array}{rcl}
\hat{\nabla}^{2}(h_{0}/\hat{f})
-\tfrac{1}{6}(\hat{F}^{0\,+})^{2}
+\tfrac{1}{12}(\hat{F}^{8\,+})^{2}+\tfrac{1}{12}(\hat{F}^{A\,+})^{2}
& & \\
& & \\
\\
+\tfrac{1}{3}g_{0}{}^{2}[8 (h_{0}/\hat{f})^{2}-(h_{8}/\hat{f})^{2}]
+\tfrac{1}{\sqrt{3}}g_{0}\hat{J}\cdot d\hat{\omega}
& = &
0\, ,
\end{array}
\end{equation}

\noindent
while the only non-trivial components of Eq.~(\ref{eq:susy_4-2-2}) ($x=A,8$)
take the form

\begin{eqnarray}
\label{eq:selfd_dotprod08}
C_{AJK}\hat{F}^{J}\cdot \hat\star\hat{F}^{K} \propto F^{A+}\cdot
(\hat{F}^{0+}-\sqrt{2}\hat{F}^{8+})
& = & 
0\, ,
\\
& & \nonumber \\
\label{eq:Max8}
\hat{\nabla}^{2}(h_{8}/\hat{f})+
\tfrac{1}{6}\hat{F}^{0\,+}\cdot \hat{F}^{8\,+}
+\tfrac{1}{6\sqrt{2}}(\hat{F}^{8\,+})^{2}
-\tfrac{1}{6\sqrt{2}}(\hat{F}^{A\,+})^{2}
& & \nonumber \\
& & \nonumber \\
+\tfrac{1}{3}g_{0}^{2}[4 (h_{0}/\hat{f})(h_{8}/\hat{f})
-\sqrt{2} (h_{8}/\hat{f})^{2}]
& = & 
0\, .
\end{eqnarray}

If one does not want to put additional constraints on the non-Abelian field strengths $\hat{F}^{A}$,
Eq.~(\ref{eq:selfd_dotprod08}) implies

\begin{equation}
\hat{F}^{8+} = \tfrac{1}{\sqrt{2}} \hat{F}^{0+}\,,
\end{equation}

\noindent
and the closure of $\hat{F}^{0},\hat{F}^{8}$, and $\hat{J}$ together with
Eq.~(\ref{eq:F0-F8-}), leads to

\begin{equation}
d( \sqrt{2}h_{0}/\hat{f}+h_{8}/\hat{f} )\wedge \hat{J}=0\, ,
\,\,\,\,\,
\Rightarrow
\,\,\,\,\,
h_{8}=\sqrt{2}( \alpha \hat{f}-h_{0})\,,
\end{equation}

\noindent
for some constant $\alpha$.  Substituting in Eq.~(\ref{eq:cubic_constr08}) we
can solve this constraint, finding these expressions for $h_{0}/\hat{f}$ and
$h_{8}/\hat{f}$ in terms of $\hat{f}$:

\begin{equation}
 h_{0}/\hat{f}
=
\tfrac{1}{3}\alpha\left( 2+\frac{1}{(\alpha \hat{f})^{3}} \right)\, ,
\hspace{1cm}
h_{8}/\hat{f}
=
\tfrac{\sqrt{2}}{3}\alpha\left(1-\frac{1}{(\alpha \hat{f})^{3}} \right)\, ,
\end{equation}

\noindent
and using all these results in Eq.~(\ref{eq:hF+08}), we get

\begin{equation}
\label{eq:domega+}
(d\hat{\omega})^{+}=\tfrac{\sqrt{3}}{2}\alpha \hat{F}^{0+}\, .
\end{equation}

On the other hand, adding Eqs.~(\ref{eq:Max0}) and (\ref{eq:Max8}) divided by
$\sqrt{2}$ gives

\begin{equation}
\label{eq:domega-}
 \hat{J}\cdot d\hat{\omega}
=
-\tfrac{4}{\sqrt{3}}g_{0}\alpha^{2}\left(1+\frac{1}{(\alpha\hat{f})^{3}}\right)\, ,
\end{equation}

\noindent
and expanding the anti-selfdual part of $d\hat{\omega}$ in the basis of
anti-selfdual 2-forms $\hat{\Phi}^{1,2,3}$ ($\hat{J}=\hat{\Phi}^{1}$) we find
that  

\begin{equation}
(d\hat{\omega})^{-}
=
-\tfrac{1}{\sqrt{3}}g_{0}\alpha^{2}
\left(1+\frac{1}{(\alpha\hat{f})^{3}}\right)\hat{J} 
+\Omega_{(2)} \hat{\Phi}^{2}
+\Omega_{(3)} \hat{\Phi}^{3}
\equiv
-\tfrac{1}{\sqrt{3}}g_{0}\alpha^{2}
\left(1+\frac{1}{(\alpha\hat{f})^{3}}\right)\hat{J} 
+d\tilde{\omega} \, .  
\end{equation}

In order to make progress, we assume again that the K\"ahler base space admits
a holomorphic isometry and therefore it can be put in the canonical form
Eq.~(\ref{eq:Kahlerisometric}). We also assume that $H=H(\varrho)$
($\varrho=x^{2}$) and $W^{2}=\Psi(\varrho)\Phi_{(k)}(x^{1},x^{3})$, which
leads to the relation Eq.~(\ref{eq:HPsi}) between $H(\varrho)$ and
$\Psi(\varrho)$. Here we are going to consider the two possible values of
$\epsilon=0,1$.

From Eq.~(\ref{eq:R=dP}), and using the results in Appendix B of
Ref.~\cite{Chimento:2016mmd}, one gets

\begin{align}
\label{eq:F0+ex}
 \hat{F}^{0+}
&
=
-\frac{1}{4 g_{0} \varrho^{\epsilon}}
\left(\Psi''-2\epsilon\frac{\Psi'}{\varrho}+2 k \right)
\left[ \left( dz+\chi_{(k)} \right)\wedge d\varrho+\varrho^{\epsilon}
  \Phi_{(k)} dx^{3}\wedge dx^{1} \right]\,,
 \\
& \nonumber\\
\label{eq:F0-ex}
\hat{F}^{0-}
& =
-\frac{1}{4 g_{0} \varrho^{\epsilon}}
\left( \Psi''-2 k \right)
\left[ \left( dz+\chi_{(k)} \right)\wedge d\varrho-\varrho^{\epsilon}
  \Phi_{(k)} dx^{3}\wedge dx^{1} \right]
=
-\frac{\hat{J}}{4 g_{0}  \varrho^{\epsilon}}\left( \Psi''-2 k \right)\, ,
\end{align}

\noindent
and comparing with the expression for $\hat{F}^{0-}$ in Eq.~(\ref{eq:F0-F8-}) one has

\begin{equation}
\label{eq:hatf}
\frac{1}{(\alpha \hat{f})^{3}}
=
\frac{3}{8 g_{0}^{2} \alpha\varrho^{\epsilon}}\left( \Psi''-2 k \right)-2\, ,
\end{equation}

\noindent
so that

\begin{align}
h_{0}/\hat{f}
&
=
\frac{\Psi''-2 k}{8 g_{0}^{2} \varrho^{\epsilon}}\, ,
\\
& \nonumber \\
h_{8}/\hat{f}
&
=
\sqrt{2}\left( \alpha- \frac{\Psi''-2 k}{8 g_{0}^{2} \varrho^{\epsilon}}\right)\, .
\end{align}

Finally, substituting everything in Eq.~(\ref{eq:Max8}) gives a fourth order
differential equation for the function $\Psi$

\begin{multline}
\label{eq:psi_diff_eq}
3\epsilon (\Psi')^{2}+6 k \varrho^{2} \Psi''
-3 \varrho\Psi' (4\epsilon k+\varrho\Psi''')
+3\Psi(4\epsilon k-2 \epsilon \Psi''+2 \epsilon \varrho \Psi'''-\varrho^{2} \Psi'''')\\
\\
-2 g_{0}^{2}\varrho^{2+\epsilon}
\left[ 
\varrho^{\epsilon} 
(\hat{F}^{A}\cdot \hat{F}^{A}+8 \alpha^{2}  g_{0}^{2})
-4\alpha (\Psi''-2k) 
\right]=0\, ,
\end{multline}

\noindent
which can only be solved if we first find a selfdual $\mathrm{SU}(2)$
instanton on the K\"ahler base space $\hat{F}^{A-}=0$. Since we only know
solutions of this kind for $k=1$ (see Ref.~\cite{Chimento:2017tsv} and the
discussion in the previous section), we will now carry a case by case analysis
of the possible solutions for different values of $\epsilon$ setting $k=1$ and
taking into account that, for any given $\Psi(\varrho)$, there are in general
two selfdual $\mathrm{SU}(2)$ instanton solutions: a ``universal'' solution
and a ``constrained'' solution.

Let us start by considering the $\epsilon=1$ case.

\subsubsection{The $\epsilon=1$ case}

Following Ref.~\cite{Chimento:2017tsv}, for this case the ``universal''
instanton solution is, irrespectively of the form of $\Psi(\varrho)$, given by

\begin{equation}
\hat{A}^{A}
=
-\frac{1}{g\varrho^{2}}
\left[
( \tfrac{1}{2}g\beta-\varrho )y^{A}
\left( dz+\chi_{(1)}\right)+\varepsilon^{A}{}_{BC}y^{B}dy^{C}  
\right]\, ,
 \quad
\Rightarrow
\quad 
\hat{F}^{A}\cdot \hat{F}^{A} = \beta^{2}/\varrho^{4}\,,
\end{equation}

\noindent
where $\beta$ is an arbitrary constant and $y^{A}$ are Cartesian coordinates
related to $\varrho$ by $y^{A}y^{A}=\varrho^{2}$.

Then, using this instanton solution in Eq.~(\ref{eq:psi_diff_eq}) and assuming
$\Psi(\varrho)$ to be a polynomial, we get  two distinct solutions, both
for $\Psi$ of order 3, $\Psi=\sum_{i=0}^{3} \psi_{i}\varrho^{i}$:

\begin{eqnarray}
& \mathbf{(u1)} &
\psi_{3}=\tfrac{4}{3} \alpha g_{0}^{2}\, ,
\,\,\,\,\,\,
\text{and}
\,\,\,\,\,\,
\beta^{2}=\frac{3}{2 g_{0}^{2}}\left[\psi_{1}^{2}-4\psi_{0}(\psi_2-1)\right]\,,
\\
& & \nonumber \\
& \mathbf{(u2)} &
\,\,\,\,\,
\psi_{3}=\tfrac{4}{9} \alpha g_{0}^{2}\, , 
\,\,\,\,\,\,
\psi_{2}=1\,,
\,\,\,\,\,\,
\text{and}
\,\,\,\,\,\,
\beta^{2}=\tfrac{3}{2} \frac{\psi_{1}^{2}}{g_{0}^{2}}\, ,
\end{eqnarray}

The ``constrained'' instanton solution is obtained by assuming from the start
that $\Psi=\sum_{i=0}^{3} \psi_{i}\varrho^{i}$, and is characterized by the
function

\begin{equation}
K
=
\frac{\psi_{3}}{2}\varrho^{2}
+\frac{\psi_{2}-1}{3}\varrho+\frac{1}{18\psi_{3}}\left[9\psi_{1}\psi_{3}
-2(\psi_{2}+2)(\psi_{2}-1)\right]\, ,
\end{equation}

\noindent
with the constraint

\begin{equation}
\label{eq:constraint}
\psi_{0}
=
\frac{\psi_{2}+2}{27\psi_{3}^{2}}
\left[9\psi_{1}\psi_{3}-2(\psi_{2}+2)(\psi_{2}-1)  \right]\, ,
\end{equation}

\noindent
and leads to the instanton number density

\begin{equation}
\hat{F}^{A}\cdot \hat{F}^{A} 
=
\frac{4}{g^{2}\varrho^{2}}
\left[\left(K'-\frac{K}{\varrho}\right)^{2}+\frac{2}{\Psi}K^{2}(K'+1)\right]\, .
\end{equation}

Substituting this expression into Eq.~(\ref{eq:psi_diff_eq}) and taking into
account the above constraints between the coefficients of the polynomial
$\Psi$ we find that,
depending on the relative value of the two coupling constants which we denote
with the parameter $\xi$

\begin{equation}
\label{eq:xidef}
\xi \equiv g/g_{0}\, .
\end{equation}

\noindent
the differential equation admits two different solutions:

\begin{eqnarray}
\label{eq:regular_sol}
& \mathbf{(c1)} &
\xi^{2}\neq 2/3\, ,
\,\,\,\,\,
\psi_{3}
=
\frac{4\alpha g_{0}^{2}}{3} \frac{6\pm\sqrt{9-6 \xi^{-2}}}{9+2 \xi^{-2}}\, ,
\,\,\,\,\,
\psi_{2}=1\, ,
\,\,\,\,\,
\psi_{1}=\psi_{0}=0\, ,
\\
& & \nonumber \\
& \mathbf{(c2)} &
\xi^{2}=2/3\, ,
\,\,\,\,\,
\psi_{3}=\frac{2\alpha g_{0}^{2}}{3}\, ,
\,\,\,\,\,
\psi_{2}=1\, ,
\,\,\,\,\,
\psi_{0}=\frac{3 \psi_{1}}{2\alpha g_{0}^{2}}\, .
\end{eqnarray}

\noindent
In this last case $\psi_{1}$ remains undetermined.

In the four cases $\mathbf{u1,u2,c1,c2}$,

\begin{equation}
\hat{\omega} 
=
\frac{\sqrt{3}\alpha}{4 g_{0}}\left\{2(\psi_{2}-1)\chi_{(1)}+\left[\frac{\psi_{1}}{\varrho}+(  3\psi_{3}-\tfrac{4}{3} g_{0}^{2}\alpha)\varrho  \right](dz+\chi_{(1)})  \right\}
 +\tilde{\omega}\,,
\end{equation}

\noindent
where we remind the reader the definition $d\tilde{\omega}\equiv\Omega_{2}
\hat{\Phi}^{(2)}+\Omega_{3} \hat{\Phi}^{(3)}$.\footnote{Under the assumption
  that the components of $\hat{\omega}$ are independent of $z$, closure
  implies
\begin{equation}
d\tilde{\omega}
=
(dz+\chi_{(k)})\wedge\mathfrak{Im}[\mathcal{H}(\zeta)d\zeta]
  +\frac{\varrho
    d\varrho}{\Psi}\wedge\mathfrak{Re}[\mathcal{H}(\zeta)d\zeta]\, ,
\end{equation}
where $\mathcal{H}$ is an arbitrary holomorphic function and $\zeta=x^{1}+i
x^{3}$.}  With a constant shift in the time coordinate $t$ it is possible to
bring $\hat{\omega}$ to the simpler form

\begin{equation}
\hat{\omega} 
=
\frac{\sqrt{3}\alpha}{4 g_{0}\varrho}
\left[
\psi_{1}+2(\psi_{2}-1)\varrho+(  3\psi_{3}-\frac43 g_{0}^{2}\alpha)\varrho^{2}  
\right](dz+\chi_{(1)})
+\tilde{\omega}\, .
\end{equation}

Also, in the four cases the function $\hat{f}$ is given by

\begin{equation}
(\alpha \hat{f})^{-3}
=
\frac{3}{4 g_{0}^{2}\alpha\varrho}(3\psi_{3}\varrho+\psi_{2}-1)-2\, ,
\end{equation}

\noindent
and this ends the determination of all the building blocks of the solutions,
which we now have to analyze. 

From now on we take $\tilde{\omega}=0$ for the sake of simplicity.  Then, 
it is possible to set the constant $\alpha$ to an arbitrary value $\alpha/\delta$
with the rescaling $\varrho\to\varrho/\delta$, $t\to \delta t$, 
$\psi_{3}\to \delta \psi_{3}$, $\psi_{1}\to\psi_{1}/\delta$, 
$\psi_{0}\to \psi_{0}/\delta^{2}$. This will allow us later to normalize 
the solution in the most convenient way.

Every solution of this form presents a (naked) curvature singularity in
$\varrho=0$, except for $\psi_{2}=1$, $\psi_{1}=\psi_{0}=0$, in which case the
curvature scalars $R$, $R_{ab}R^{ab}$ and $R_{abcd}R^{abcd}$ are
constant. This is the case for the solution $\mathbf{c1}$ in
(\ref{eq:regular_sol}). The solutions $\mathbf{u2}$ and $\mathbf{c2}$ give a metric with the
wrong signature. This leaves us with the only meaningful possibilities:

\begin{description}
\item[$\mathbf{u1}$] singular at $\varrho=0$ except for $\psi_{2}=1$,
  $\psi_{1}=\psi_{0}=0$, in which case $\beta=0$, the matter fields are
  trivial and the solution is just AdS$_{5}$.

\item[$\mathbf{c1}$] regular. Defining the parameter 

\begin{equation}
 \gamma^{-1}=2\mp\sqrt{1-\tfrac{2}{3} \xi^{-2}}\, ,
\,\,\,\,\,
\Rightarrow
\,\,\,\,\,
\psi_{3}=\tfrac{4}{3}\alpha \gamma g_{0}^{2}\, ,
\,\,\,\,\,
\text{and}
\,\,\,\,\,
\hat{f}^{-1}= \alpha(3\gamma -2)^{1/3}\, ,
\end{equation}

\noindent
it's easy to see that for the metric to have the right signature one has to take 
the upper sign in the definition of $\gamma$ and to impose $\gamma > 2/3$, or 
equivalently $\xi^2 > 8/9$.

\noindent
Then it is possible to use the rescaling mentioned above to adjust the integration 
constant $\alpha$ so that $\hat{f}=1$ ($\alpha=(3\gamma -2)^{-1/3}$) and to define 
$\tilde{g}$ and $\lambda$ by

\begin{equation}
\tilde{g}_{0}^{2}
\equiv 
\alpha \gamma g_{0}^{2} 
= 
\frac{\gamma}{(3\gamma  -2)^{1/3}}g_{0}^{2}\, ,
\,\,\,\,\,
\text{and}
\,\,\,\,\,
1+\lambda \equiv \frac{3\gamma-1}{2(3\gamma^{2}-2\gamma)^{1/2}}\, ,   
\end{equation}

\noindent
so that the metric takes the form of that in the solution
Eq.~(\ref{eq:quasiads5k1}) with the replacement $g_{0}\to \tilde{g}_{0}$. This
happens because the scalar potential for these solutions also takes the same
value with the replacement of $g_{0}\to \tilde{g}_{0}$,
$V(\phi)=-4\tilde{g}_{0}$. The remaining non-vanishing fields of the solution
are

\begin{eqnarray}
F^{0}
& = &
\tfrac{1}{\sqrt{2}} F^{8}
=
-g_{0}\frac{\gamma-1}{(3\gamma-2)^{4/3}}\hat{J}\, ,
\\
& & \nonumber \\ 
\phi 
& \equiv &
h_{8}/h_{0}
=\sqrt{2}( \gamma^{-1}-1 )\, ,
\\
& & \nonumber \\ 
A^{A}
& = &
\hat{A}^{A}
=
-\frac{2 \tilde{g}_{0}^{2}}{3 g}y^{A} (dz+\cos{\theta} d\varphi) 
+\frac{(1+ \tfrac{4}{3} \tilde{g}_{0}^{2} \varrho)^{1/2}-1}{g\varrho^{2}}
\epsilon^{A}{}_{BC}y^{B}dy^{C}\, .
\end{eqnarray}

The instanton number density is given by 

\begin{equation}
\hat{F}^{A}\cdot \hat{F}^{A} = \tfrac{16}{3}\frac{\tilde{g}_{0}^{4}}{g^{2}}\, .  
\end{equation}

In the limit $g\to\infty$ for fixed $g_{0}$ then $\gamma$ goes to $1$ and the
above solution reduces to AdS$_5$.


\end{description}

\subsubsection{The $\epsilon=0$ case}

For $\epsilon=0$ Eq.~(\ref{eq:psi_diff_eq}) takes the much simpler form 

\begin{equation}
\label{eq:psi_diff_eqe=0}
\Psi' \Psi''' +\Psi \Psi''''-2\Psi''
+\tfrac{2}{3} g_{0}^{2}
\left[ 
\hat{F}^{A}\cdot \hat{F}^{A}+8 \alpha^{2}  g_{0}^{2}
-4\alpha (\Psi''-2) 
\right]=0\, .
\end{equation}
This equation admits no solution for the constrained instanton solution.
Let us then consider the universal solution, for which the instanton number
density is always given by 

\begin{equation}
\hat{F}^{A}\cdot \hat{F}^{A}=\frac{4}{g^{2}}\, .
\end{equation}

\noindent
and Eq.~(\ref{eq:psi_diff_eqe=0}) becomes

\begin{equation}
\label{eq:psi_diff_eqe=0universal}
\Psi' \Psi''' +\Psi \Psi''''-2\Psi''
-\tfrac{8}{3} \alpha g_{0}^{2} (\Psi''-2 -2\alpha g^{2}_{0})
+\tfrac{8}{3}(g_{0}/g)^{2}
=0\, .
\end{equation}

If, as usual, we assume $\Psi$ to be a polynomial in $\varrho$, from
Eq.~(\ref{eq:psi_diff_eqe=0universal}) we find that it is at most of second order,
$\Psi=\sum_{i=0}^{2} \psi_{i}\varrho^{i}$. There are two possibilities to solve the differential equation (\ref{eq:psi_diff_eqe=0universal}):

\begin{enumerate}

\item $\mathbf{u1}$ If $a\equiv \alpha g_{0}^{2}\neq -\frac{3}{4}$,
  Eq.~(\ref{eq:psi_diff_eqe=0universal}) is satisfied for

\begin{equation}
\label{eq:eps0psi2}
\psi_{2}=\frac{2\xi^{-2}+4a (1+a)}{3+4a}\, ,
\end{equation}

\noindent
where $\xi$. $\psi_{0}$ and $\psi_{1}$ are left unconstrained.

\item $\mathbf{u2}$ If $a = -\frac{3}{4}$
  Eq.~(\ref{eq:psi_diff_eqe=0universal}) only admits a solution for a specific
  value of $\xi\equiv g/g_{0}$:

\begin{equation}
\label{eq:eps0fixedxi}
\xi^{-2}=\frac{3}{8}\,.
\end{equation}

\noindent
In this case the polynomial $\Psi$ is not constrained by these equations.

\end{enumerate}

In both cases we can, again, impose $\tilde{\omega}=0$ for simplicity, 
use Eqs.~(\ref{eq:domega+}), (\ref{eq:domega-}), (\ref{eq:F0+ex}), integrate
to obtain $\hat{\omega}$, and write the five-dimensional metric as

\begin{equation}
ds^{2}
=
\hat{f}^{2}\left(dt - c_{1}\varrho dz-c_{2} \cos{\theta} d\varphi\right)^{2}
-\hat{f}^{-1} \left[\Psi dz^{2}+\frac{d\varrho^{2}}{\Psi}+d\Omega^{2}_{(2,1)}\right]\ ,
\end{equation}

\noindent
where the constants $c_{1,2}$ are given by 

\begin{equation}
c_{1}
=
\frac{a}{ \sqrt{3}g_{0}^{3}}
\left(a-\tfrac{3}{2}\psi_{2}\right)\, , 
\hspace{1cm}
c_{2}
=
\frac{a}{\sqrt{3}g_{0}^{3}}
\left(a+\tfrac{3}{2}\right)\, ,
\end{equation}

\noindent
and $\hat{f}$ is determined from Eq.~(\ref{eq:hatf}) to be constant:

\begin{equation}
\hat{f}^{-3}=\frac{3a^{2}\left(\psi_{2}-1\right)-8a^{3}}{4g_{0}^{6}}\, .
\end{equation}

The general structure of the metric is that of a $\mathrm{U}(1)$ fibration
over the product of 2 2-dimensional spaces: the 2-sphere and the space
parametrized by $(\varrho,z)$, which we are going to study in more detail
below.

The complete non-Abelian 1-form field and its 2-form field strength are given
by

\begin{equation}
\label{eq:nonabelianepsilon0}
A^{A}
=
\hat{A}^{A}
=
\frac{1}{g}
\left(
y^{A} dz-\frac{1}{\varrho^{2}}\epsilon^{A}{}_{BC}y^{B}dy^{C}
\right)\, ,
\hspace{.5cm}
F^{A} 
= 
\hat{F}^{A} 
=
\frac{y^{A}}{g\varrho}
\left( d\varrho\wedge dz+\sin\theta d\theta\wedge d\varphi \right)\, ,
\end{equation}

\noindent
thus, the field strength is $1/g$ times the unit vector $y^{A}/\varrho$ times
the sum of the volume forms of the 2-dimensional spaces that enter in the base
space.

The remaining fields take the form

\begin{eqnarray}
A^{0} 
& = &
\frac{-2 a/g_{0}}{3(\psi_{2}-1)-8 a}\left[(1+2 a)\varrho dz+(2
  a-\psi_{2})\cos\theta d\varphi\right]\,,
\\
& & \nonumber \\
A^{8}
& = &
-\sqrt{2} A^{0}\, ,
\\
& & \nonumber \\
\phi
& = &
\frac{g_{0}^{2}}{a\hat{f}}\,.
\end{eqnarray}

By shifting and rescaling the coordinates $\varrho$ and $z$ we can reduce the
number of independent parameters and study in more detail the possible 2- and
5-dimensional metrics that arise

\begin{itemize}

\item If $\psi_{2} \neq 0$, we can bring the base space metric to the form 

\begin{equation}
d\hat{s}^{2}
=
\frac{1}{\psi_{2}}
\left[(\varrho^{2}-\varepsilon)dz^{2}
+\frac{d\varrho^{2}}{\varrho^{2}-\varepsilon}\right]
+d\Omega^{2}_{(2,1)}\, ,
\end{equation}

\noindent
where $\varepsilon=0,\pm 1$. The Ricci scalar of the two-dimensional space
parametrized by $(\varrho,z)$ is constant and negative for any of the three
values of $\varepsilon$ and so, it is maximally symmetric. Therefore this is
the metric of the hyperbolic plane that we will denote by
$d\Omega^{2}_{(2,-1)}$\footnote{Actually, it is easy to see that a simple
  coordinate change brings the metric to the standard form of the hyperbolic
  plane in polar, Lobachevsky and Poincar\'e half-plane coordinates
  respectively for $\varepsilon=1,-1,0$.}  and the base-space metric is that
of the product of $\mathbb{H}_{2}$ with radius squared $1/\psi_{2}$ and
$\mathrm{S}^{2}$ with radius $1$

Denoting by $ \chi_{(1)}\equiv \cos{\theta}d\varphi$ the K\"ahler 1-form of
the 2-sphere and by $ \chi_{(-1)}\equiv \varrho dz$ the K\"ahler 1-form of the
hyperbolic plane,  the full five-dimensional metric (after shifting the time
coordinate) and the rest of the fields can be written as

\begin{eqnarray}
ds^{2}
& = &
\hat{f}^{2}\left(dt - \frac{c_{1}}{\psi_{2}}\chi_{(-1)}-c_{2}
\chi_{(1)}\right)^{2}
-\hat{f}^{-1} \left[\frac{1}{
    \psi_{2}}d\Omega^{2}_{(2,-1)}+d\Omega^{2}_{(2,1)}\right]\, ,
\\
& & \nonumber \\
A^{A}
& = &
\frac{1}{g \psi_{2}}\left(y^{A}
  dz-\frac{\psi_{2}}{\varrho^{2}}\epsilon^{A}{}_{BC}y^{B}dy^{C}  \right)\, ,
\\
& &\nonumber \\
F^{A} 
& = &
\frac{y^{A}}{g \varrho}
\left(
\frac{1}{\psi_{2}} d\varrho\wedge dz+\sin{\theta} d\theta\wedge d\varphi 
\right)\, ,
\\
& & \nonumber \\
A^{0} 
& = &
\frac{-2 a/g_{0}}{3(\psi_{2}-1)-8 a}\left[
\frac{(1+2 a)}{\psi_{2}}\chi_{(-1)}+(2 a-\psi_{2})\chi_{(1)}\right]\, ,
\\
& & \nonumber \\
A^{8}
& = &
-\sqrt{2} A^{0}\, ,
\\
& & \nonumber \\
\label{eq:scalar}
\phi
& = &
\frac{g_{0}^{2}}{a\hat{f}}\,.
\end{eqnarray}

This metric has the same form as the G\"odel solutions found in
\cite{Chimento:2016mmd}. It is well known that, generically, these metrics
have closed timelike curves (CTCs), but one can wonder if it is possible to
tune the parameter $a$ in such a way as to avoid them. This would demand
setting $c_{2}=0$ ($a=-3/2$) to avoid Misner string singularities or having to
compactify the time coordinate to avoid them. It is not possible to set
$c_{1}=0$ at the same time ($\psi_{2}$ must be strictly positive). Then, the
condition for absence of CTCs is

\begin{equation}
(\hat{f}c_{1})^{2}< \hat{f}^{-1}/\psi_{2}\, ,  
\end{equation}

\noindent
and with $a=-3/2$, this condition cannot be satisfied for any value of
$\psi_{2}$.

Note also that the metric only has the correct signature if $\hat{f}$ and $\psi_2$ are 
both positive, which means $a\in (-\tfrac{3}{4},0)$.

\item If $\psi_{2}=0$ and $\psi_{1}\neq 0$, we get a 5-dimensional metric of
  the form

\begin{equation}
ds^{2}
=
\hat{f}^{2}\left[dt -c_{1}\varrho dz-c_{2} \chi_{(1)}\right]^{2}
-
\hat{f}^{-1}\left[\varrho dz^{2}+\frac{d\varrho^{2}}{\varrho}+d\Omega^{2}_{(2,1)}\right]\, .
\end{equation}

\item If $\psi_{2}=\psi_{1}=0$ and $\psi_{0}> 0$,\footnote{For $\psi_{0}<0$
    one would get the wrong signature for the metric.} we get a 5-dimensional
  metric of the form

\begin{equation}
ds^{2}
=
\hat{f}^{2}\left[dt -c_{1}\varrho dz-c_{2}\chi_{(1)}\right]^{2}
-\hat{f}^{-1}\left[dz^{2}+d\varrho^{2}+d\Omega^{2}_{(2,1)}\right]\, .
\end{equation}

\end{itemize} 

The metric for the last two cases is actually the same one written in
different coordinates, and can also be written as

\begin{equation}
ds^{2}
=\hat{f}^{2}\left[dt -c_{1}\chi_{(0)}-c_{2} \chi_{(1)}\right]^{2}
-\hat{f}^{-1}\left[d\Omega^{2}_{(2,0)}+d\Omega^{2}_{(2,1)}\right]\,,
\end{equation}

\noindent
where $d\Omega^{2}_{(2,0)}$ is the metric of the 2-dimensional Euclidean space
and $\chi_{(0)}$ its K\"ahler 1-form and with

\begin{equation}
c_{1}
= 
\frac{a^{2}}{\sqrt{3}g_{0}{}^{3}}\, ,
\hspace{1cm}
c_{2}
=
\frac{a}{\sqrt{3}g_{0}^{3}}
\left(a+\tfrac{3}{2}\right)\, ,
\hspace{1cm}
\hat{f}^{-3}
=
-\frac{a^{2}(3+8a)}{4g_{0}^{6}}\, .
\end{equation}

The non-Abelian fields are given by the general expressions
Eqs.~(\ref{eq:nonabelianepsilon0}) and the Abelian ones by

\begin{eqnarray}
A^{0} 
& = & 
\frac{2 a/g_{0}}{3+8 a}\left[(1+2 a)\chi_{(0)}+2 a\chi_{(1)}\right]\, ,
\\
& & \nonumber \\
A^{8}
& = &
-\sqrt{2} A^{0}\, ,
\end{eqnarray}

\noindent
while the constant scalar field is still given by Eq.~(\ref{eq:scalar}).

Observe that, in this case, $a$ is not a free parameter, since $\psi_{2}=0$
implies from Eqs.~(\ref{eq:eps0psi2}) and (\ref{eq:eps0fixedxi})

\begin{equation}
\label{eq:aparam}
 a=\frac{-1\pm\sqrt{1-2\xi^{-2}}}{2}\, .
\end{equation}
For this condition to make sense one must of course impose $\xi^2\geq 2$.

\section{Embedding in half-maximal $d=5,$
  $\mathrm{SU}(2)\times\mathrm{U}(1)$-gauged supergravity}
\label{sec-n4embedding}

Uplifting the solutions of the cosmological gauged $\mathbb{C}$ magic model to
10 dimensions presents severe practical difficulties, starting with the
embedding of the solutions we have obtained in $\mathrm{SO}(6)$-gauged $d=5$
maximal supergravity which requires a definite relation between the
$\mathrm{U}(1)_{\rm R}$ and $\mathrm{SU}(2)$ coupling constants which is not
readily available in the literature. Then, one would have to face the problem
of uplifting the solution to 10 dimensions.

There are not many reduction anstazs that lead from 10-dimensional
supergravities to gauged 5-dimensional supergravities and which permit the
automatic uplifting of the 5-dimensional solutions, especially if one is
interested in a particular gauge group. There is, however, a reduction ansatz
from $\mathcal{N}=2B,d=10$ supergravity to gauged, half-maximal $d=5$
supergravity with, precisely, the gauge group
$\mathrm{SU}(2)\times\mathrm{U}(1)$ \cite{Lu:1999bw}.\footnote{We thank
  O.~Varela for pointing this reference to us.} Since the ansatz corresponds
to compactification on a 5-sphere it is natural to expect that the two gauge
coupling constants are not independent and, as we shall see, in fact one gets, in our
conventions, $\xi^{2}=(g/g_{0})^{2}=2/3$.

We would like to embed our solutions in this 5-dimensional theory in order to
be able to uplift them to 10 dimensions but the relation $\xi^{2}=2/3$ will
only allow us to uplift some of them. It is, by no means, guaranteed that such
an embedding is possible but we are going to show that indeed it is for some
solutions of the cosmological $\mathbb{C}$ gauged magic model that include
several of those we have constructed in the previous section. More
specifically, we are going to show that the consistently truncated equations
of motion of the cosmological $\mathbb{C}$ gauged magic model (for a truncation that
includes the solutions we have constructed) coincide with the consistently
truncated equations of motion of $\mathrm{SU}(2)\times\mathrm{U}(1)$-gauged,
half-maximal $d=5$ supergravity.

\subsection{Truncated equations of motion}
\label{sec-truncated}

Let us consider the equations of motion of the cosmological $\mathbb{C}$
gauged magic model (\ref{eq:eom1})-(\ref{eq:eom3}) (where we have replaced the
generic objects $a_{IJ},C_{IJK},g_{xy},k_{I}{}^{x}$ by their values for this
particular model, evidently). If we set $h_{1,\ldots,7}=h^{1,\ldots,7}=0$ and
$A^{4,\ldots,7}=0$, and define

\begin{equation}
 X\equiv(h_{0}+\tfrac{1}{\sqrt{2}}h_{8})^{-1}
\quad
\Rightarrow\quad 
h_{0}-\sqrt{2}h_{8}=X^{2}\, ,
\quad 
\mathcal{H}\equiv F^{0}+\tfrac{1}{\sqrt{2}}F^{8}\, ,
\quad 
\mathcal{G}\equiv F^{0}-\sqrt{2}F^{8}\, ,
\end{equation}

\noindent
the equations of motion reduce to\footnote{We have subtracted the trace of
  the Einstein equation.}

\begin{eqnarray}
R_{\mu\nu}
-\tfrac{1}{6} X^{4} 
(\mathcal{G}_{\mu}{}^{\rho}\mathcal{G}_{\nu\rho}
-\tfrac{1}{6} g_{\mu\nu}\mathcal{G}^{\rho\sigma}\mathcal{G}_{\rho\sigma})
-\tfrac{1}{3} X^{-2} ( \mathcal{H}_{\mu}{}^{\rho}\mathcal{H}_{\nu\rho}
-\tfrac{1}{6} g_{\mu\nu}\mathcal{H}^{\rho\sigma}\mathcal{H}_{\rho\sigma} )
& & \nonumber \\
& & \nonumber \\
-\tfrac{1}{2} X^{-2} 
( F^{A}{}_{\mu}{}^{\rho} F^{A}{}_{\nu\rho}
-\tfrac{1}{6} g_{\mu\nu}F^{A\rho\sigma}F^{A}{}_{\rho\sigma} )
& & \nonumber \\
& & \nonumber \\
 +3\,\partial_{\mu}\log{X}\,\partial_\nu\log{X} 
+\tfrac{4}{9}g_{0}^{2} g_{\mu\nu}(X^{2}+2X^{-1})
& = & 
0\, ,
\\
& & \nonumber \\
\nabla^{2}\log{X}
-\tfrac{1}{12} X^{-2} F^{A}\cdot F^{A}
-\tfrac{1}{18} X^{-2} \mathcal{H}\cdot \mathcal{H}
+\tfrac{1}{18} X^{4} \mathcal{G}\cdot \mathcal{G}
-\tfrac{4}{9} g_{0}^{2} (X^{2}-X^{-1})
& = &
0\, ,
\\
& & \nonumber \\
\label{eq:FAHconstraint}
F^{A}\cdot\mathcal{H}
& = & 
0\, ,
\\
& & \nonumber \\
\nabla_{\nu}(X^{-2}\mathcal{H}^{\nu\mu})
+\tfrac{1}{4\sqrt{3}}\frac{\epsilon^{\mu\nu\rho\sigma\alpha}}{\sqrt{g}}
\mathcal{H}_{\nu\rho}\mathcal{G}_{\sigma\alpha}
& = & 0\, ,
\\
& & \nonumber \\
\nabla_{\nu}(X^{4} \mathcal{G}^{\nu\mu})
+\tfrac{1}{4\sqrt{3}}\frac{\epsilon^{\mu\nu\rho\sigma\alpha}}{\sqrt{g}}
\mathcal{H}_{\nu\rho}\mathcal{H}_{\sigma\alpha}
-\tfrac{\sqrt{3}}{8}\frac{\epsilon^{\mu\nu\rho\sigma\alpha}}{\sqrt{g}}F^{A}{}_{\nu\rho}F^{A}{}_{\sigma\alpha}
& = &
0\, ,
\\
& & \nonumber \\
 \mathfrak{D}_{\nu}(X^{-2}F^{A\nu\mu})
-\tfrac{1}{4\sqrt{3}}\frac{\epsilon^{\mu\nu\rho\sigma\alpha}}{\sqrt{g}}
F^{A}{}_{\nu\rho}\mathcal{G}_{\sigma\alpha}
& = & 
0\, .
\end{eqnarray}

The constraint Eq.~(\ref{eq:FAHconstraint}) can be solved preserving the
non-Abelian gauge fields by setting $\mathcal{H}=0$. This also solves the
equation for $\mathcal{H}$, leaving no further constraints. The resulting
equations are

\begin{eqnarray}
R_{\mu\nu}
-\tfrac{1}{6} X^{4} 
(\mathcal{G}_{\mu}{}^{\rho}\mathcal{G}_{\nu\rho}
-\tfrac{1}{6} g_{\mu\nu}\mathcal{G}^{\rho\sigma}\mathcal{G}_{\rho\sigma})
\hspace{3cm}~
& & \nonumber \\
& & \nonumber \\
-\tfrac{1}{2} X^{-2} 
( F^{A}{}_{\mu}{}^{\rho} F^{A}{}_{\nu\rho}
-\tfrac{1}{6} g_{\mu\nu}F^{A\rho\sigma}F^{A}{}_{\rho\sigma} )
\hspace{1.5cm}~
& & \nonumber \\
& & \nonumber \\
 +3\,\partial_{\mu}\log{X}\,\partial_\nu\log{X} 
+\tfrac{4}{9}g_{0}^{2} g_{\mu\nu}(X^{2}+2X^{-1})
& = & 
0\, ,
\\
& & \nonumber \\
\nabla^{2}\log{X}
-\tfrac{1}{12} X^{-2} F^{A}\cdot F^{A}
+\tfrac{1}{18} X^{4} \mathcal{G}\cdot \mathcal{G}
-\tfrac{4}{9} g_{0}^{2} (X^{2}-X^{-1})
& = &
0\, ,
\\
& & \nonumber \\
\nabla_{\nu}(X^{4} \mathcal{G}^{\nu\mu})
-\tfrac{\sqrt{3}}{8}\frac{\epsilon^{\mu\nu\rho\sigma\alpha}}{\sqrt{g}}F^{A}{}_{\nu\rho}F^{A}{}_{\sigma\alpha}
& = &
0\, ,
\\
& & \nonumber \\
 \mathfrak{D}_{\nu}(X^{-2}F^{A\nu\mu})
-\tfrac{1}{4\sqrt{3}}\frac{\epsilon^{\mu\nu\rho\sigma\alpha}}{\sqrt{g}}
F^{A}{}_{\nu\rho}\mathcal{G}_{\sigma\alpha}
& = & 
0\, ,
\end{eqnarray}

\noindent
which, if $\xi^{2}=(g/g_{0})^{2}=2/3$ and after a rescaling of $\mathcal{G}$,
are identical to those obtained in Ref.~\cite{Lu:1999bw} when the tensor
fields in the latter are set to zero.

Since we have used the constraint $\mathcal{H}=0$ in the construction of our
solutions, we can, in principle, embed all of them in
$\mathrm{SU}(2)\times\mathrm{U}(1)$-gauged, half-maximal $d=5$ supergravity
and, then, using the dimensional reduction ansatz in Ref.~\cite{Lu:1999bw},
uplift them to solutions of $\mathcal{N}=2B,d=10$ supergravity.  However as we
have seen only some of them are compatible with the constraint $\xi^{2}=2/3$,
namely the solutions we have called $\mathbf{u1}$ in the two cases
$\epsilon=1,0$, and for $\epsilon=0$ only the subcase $\psi_{2}\neq 0$, since
otherwise it would require $\xi^{2} \geq 2$.  These present some undesirable
characteristics (a naked singularity for $\epsilon=1$ and closed timelike
curves for $\epsilon=0$) which are also present in the uplifted 10-dimensional
solutions.

\section{Conclusions}
\label{sec-conclusions}

By exploiting the supersymmetric solution-generating techniques developed over
the years we have managed to find some of the simplest non-Abelian solutions
of two models of ``cosmological, $\mathrm{SU}(2)$-gauged,''
$\mathcal{N}=1,d=5$ supergravity coupled to vector multiplets where the term
``cosmological'' refers to an additional $\mathrm{U}(1)_{\rm R}$ gauging that
gives rise to a non-vanishing scalar potential. The non-Abelian gauge field
configurations in these solutions is, by construction, that of a self-dual
instanton over a 4-dimensional K\"ahler manifold admitting a holomorphic
isometry.

We have found a solution that occurs with the same metric
Eq.~(\ref{eq:quasiads5k1}) and slightly different matter fields in both
models. This is an interesting supersymmetric 1-parameter deformation of
$\mathrm{AdS}_{5}$ which, as opposed to the deformation found in
Ref.~\cite{Gauntlett:2003fk} and studied in Ref.~\cite{Gauntlett:2004cm}, is
not asymptotically-$\mathrm{AdS}_{5}$.  It does not have a holographic screen
in the $\varrho\rightarrow \infty$ either because in this limit it is not
conformal to any regular metric. In the most obvious frame, all the components
of the Riemann tensor of this metric are constant, which implies that all its
curvature invariants are constant. It might be a homogeneous Riemannian space,
though, although we have not checked completely this possibility.

The rest of the solutions that we have found fall in two types: those which
may asymptote to $\mathrm{AdS}_{5}$ but have naked singularities at
$\varrho=0$, and those which are generalizations of the G\"odel-like solutions
found in Ref.~\cite{Chimento:2016mmd} in the context of pure cosmological
supergravity and whose metrics are timelike $\mathrm{U}(1)$ fibrations over
products of 2-dimensional maximally symmetric spaces. All of them seem to have
closed timelike curves.

Our second goal was to study the possible embedding of the solutions of the
cosmological, gauged $\mathbb{C}$ magical model in String Theory via maximal
or half-maximal gauged $d=5$ supergravity. The embedding in maximal
supergravity is only possible for the relation between the $\mathrm{U}(1)_{\rm
  R}$ and $\mathrm{SU}(3)$ gauge coupling constants $g_{0}$ and $g$, which
follows from the breaking of the $\mathrm{SO}(6)$ gauge group. Finding this
relation is a very complicated problem whose solution needs a precise
knowledge of the relation between the fields used in the formalism of
$\mathcal{N}=1,d=5$ theories and those of the maximal supergravity, which is
not available. This knowledge is also needed for actual embedding and,
therefore, although it is guaranteed that some of the solutions found can be
embedded and uplifted to 10 dimensions, the embedding and uplifting cannot be
realized in practice.

The embedding in the $\mathrm{SU}(2)\times \mathrm{U}(1)$-gauged $d=5$
half-maximal supergravity of Ref.~\cite{Lu:1999bw} also requires a precise
relation between the coupling constants, but in this case this relation is
known and also satisfied by some of the solutions, although they are the
singular ones and they remain singular after uplifting them to 10 dimensions.

The difficulties in uplifting the solutions to 10 dimensions leave us without
an interpretation of the fields in terms of branes although the regular
solution Eq.~(\ref{eq:quasiads5k1}) seems to be a generalization of the
gravitating Yang-Mills instanton of Ref.~\cite{Cano:2017sqy} since, also in
this case, in the two models we have studied, the graviphoton field is sourced
by the instanton number density 4-form. The model studied in
Ref.~\cite{Cano:2017sqy} can be obtained by a toroidal compactification and
truncation of 10-dimensional Heterotic Supergravity and the graviphoton is
related to the Kalb-Ramond 2-form. The solution is, therefore, a
compactification of the gauge 5-brane. In the theories that we have considered
the graviphoton gauges $\mathrm{U}(1)_{\rm R}$ via a Fayet-Iliopoulos term and
the 10-dimensional interpretation is much less transparent.

Although the balance of this work in terms of interesting solutions (especially
from the holographic point of view) may look slightly disappointing, it is
fair to say this is just the beginning of the exploration of a large, unknown, and
very complicated territory. We have put to work all the techniques developed
in the field and showed that they work in these very complicated systems. Just
as in the asymptotically-flat case, more interesting supersymmetric
non-Abelian solutions must exist and we expect to be able to find some of them
in forthcoming work.

\section*{Acknowledgments}

The authors would like to thank O.~Varela for interesting conversations and
for pointing Refs.~\cite{Lu:1999bw} and~\cite{Gauntlett:2004cm} to them. The
work of AR has been supported by a \textit{Campus Internacional de Excelencia
  UAM-CSIC} FPI grant and also by a \textit{Residencia de Estudiantes} grant.  This work has been supported in part by the
MINECO/FEDER, UE grant FPA2015-66793-P and by the Spanish Research Agency
(Agencia Estatal de Investigaci\'on) through the grant IFT Centro de
Excelencia Severo Ochoa SEV-2016-0597 which has also suppported the work of
SC.  TO wishes to thank M.M.~Fern\'andez for her permanent support.

\appendix


\end{document}